\documentclass[11pt]{article}
\usepackage[utf8]{inputenc}
\usepackage[authoryear]{natbib}
\usepackage[a4paper]{geometry}
\usepackage[title]{appendix}
\usepackage{xcolor}
\usepackage{setspace}
\usepackage{algorithm}
\usepackage{algpseudocode}
\usepackage{tikz}
\usetikzlibrary{positioning, fit, backgrounds}
\algrenewcommand\algorithmicrequire{\textbf{Input:}}
\algrenewcommand\algorithmicensure{\textbf{Output:}}
\usepackage{adjustbox}
\usepackage{booktabs}

\usepackage{amsmath}

\usepackage{amsfonts}
\usepackage{bbm}
\usepackage{graphicx}
\usepackage{booktabs, multirow}
\usepackage{makecell}
\usepackage{amsthm}
\theoremstyle{definition}

\usepackage{url}
\usepackage{comment}
\usepackage{subcaption}

\usepackage[normalem]{ulem} 
\numberwithin{equation}{section}

\title{How does academic performance affect self-efficacy? Interpretable modelling through latent academic achievement}

\author{
Sarah Lee\textsuperscript{1,2}
\and
Matias Quiroz\textsuperscript{1,2,}\textsuperscript{*}
\and
Sally Cripps\textsuperscript{1,2}
}
\date{}

\begin{document}

\maketitle

\footnotetext[1]{Human Technology Institute, University of Technology Sydney, Sydney, NSW, Australia.}
\footnotetext[2]{School of Mathematical and Physical Sciences, University of Technology Sydney, Sydney, NSW, Australia.}

\begingroup
\renewcommand{\thefootnote}{\fnsymbol{footnote}}
\footnotetext[1]{Corresponding author. Email: \texttt{quiroz.matias@gmail.com}.}
\endgroup

\begin{abstract}
There is increasing evidence of a directional relationship from academic performance to self-efficacy. We develop a Bayesian model for investigating this relationship when academic performance is measured on an ordinal scale and self-efficacy on a continuous scale. The model allows latent academic achievement to enter the self-efficacy regression as a predictor, while Bayesian variable selection identifies factors associated with either response. The resulting conditional formulation yields an interpretable regression characterisation of how latent academic achievement relates to self-efficacy. Furthermore, it enables a tailored partially collapsed Gibbs sampler that analytically integrates out the regression coefficients when updating the variable inclusion indicators. Simulation studies demonstrate that the proposed conditional formulation and tailored sampler improve sampling efficiency and variable-selection performance relative to a recent, more general joint Gaussian copula regression formulation. We apply the methodology to data from the longitudinal study of Australian children, a landmark national cohort study covering children's education, social and emotional wellbeing, health and family circumstances. The model and analysis shed light on how latent academic achievement relates to self-efficacy in Australian children, and reveal that the two outcomes differ markedly in the range of covariates associated with each outcome.

 
\end{abstract}

\section{Introduction}\label{sec:introduction}

Academic achievement and self-efficacy are two important indicators of educational development. Academic achievement is a central educational outcome and is closely linked to future educational and labour-market opportunities \citep{oecd2018equity}. Self-efficacy refers to an individual's belief in their capability to successfully perform tasks and attain desired outcomes \citep{bandura1997self}, and its importance in educational settings is discussed in \citet{pajares1996self}. A growing body of educational research suggests that academic achievement and self-efficacy are closely related. Self-efficacy has been associated with a variety of academic outcomes \citep{lent1991selfefficacy,gore2006academic}. While much of the literature has focused on the influence of self-efficacy on academic performance, increasing evidence suggests that academic performance may itself be an important antecedent of subsequent self-efficacy \citep{talsma2018believe, hwang2016relationship, burns2019}. Motivated by this evidence, we formulate our joint model in the academic performance-to-self-efficacy direction, allowing us to investigate whether academic performance remains associated with self-efficacy after accounting for a rich set of individual, family, school, health and socioeconomic characteristics. 

In this study, we use parents’ assessment of their child’s academic achievement, rather than a more objective measure such as a test-based score, because self-efficacy is theorised to arise from interpreted mastery experience rather than objective performance alone \citep{bandura1997self,WoodBandura1989Ability}. Throughout the paper, references to academic achievement and academic performance refer to this parental assessment measure. Parents' assessments are likely to reflect both the child’s prior academic performance and the evaluative context through which that performance is understood, making it more proximal to the formation of self-efficacy than a standardised test score. This interpretation is supported by a recent causal analysis submitted to the Australian House of Representatives Standing Committee on Education’s inquiry into the factors driving educational attainment,  which finds evidence that parents’ assessments of a child's academic achievement influence self-efficacy rather than the reverse \citep{CrippsGraceFrancis2026EducationalAttainment}. Understanding the factors associated with academic achievement and self-efficacy, and quantifying their relationship, is therefore of considerable interest for educational researchers and policy makers. Academic achievement is known to be shaped by a wide range of factors spanning the individual, family, school and health domains \citep{hattie2009visible}, whereas self-efficacy is theorised to be determined more proximally by interpreted mastery experience \citep{WoodBandura1989Ability, bandura1997self}; the proposed variable-selection framework is well placed to characterise this distinction empirically.

The Longitudinal Study of Australian Children (LSAC) provides a unique opportunity to investigate these questions in the Australian context \citep{sanson2002lsac}. This nationally representative longitudinal study contains rich information on children's educational outcomes, family circumstances, health, social environment and emotional well-being. In addition to providing measures of academic performance and self-efficacy, the study contains a large collection of potential explanatory variables spanning multiple aspects of children's lives. These data make it possible to investigate not only the factors associated with each outcome individually, but also the relationship between academic achievement and self-efficacy. This is particularly relevant in Australia, where educational attainment, social mobility and student well-being remain important policy concerns \citep{doe2025betterfairerschools,morris2024inequality}.

The joint analysis of academic performance and self-efficacy presents several statistical challenges. First, parents' assessment of academic performance is recorded on an ordinal scale, whereas self-efficacy is measured on a continuous scale. Standard multivariate Gaussian regression models are therefore not directly applicable. Second, academic achievement and self-efficacy are expected to be associated, suggesting that separate analyses may fail to fully utilise the available information. Third, it is plausible that some predictors influence academic achievement, some influence self-efficacy, some influence both outcomes and others influence neither. A flexible modelling framework should therefore accommodate outcome-specific variable selection while accounting for the dependence between the two responses. To address these challenges, we develop a Bayesian variable-selection approach for bivariate continuous--ordinal regression. The ordinal parents' assessment of academic performance is represented through an underlying latent Gaussian variable, while self-efficacy is modelled conditionally on latent academic achievement. This conditional formulation is motivated by recent evidence of a directional relationship from academic achievement to self-efficacy \citep{CrippsGraceFrancis2026EducationalAttainment}. Bayesian variable selection is incorporated separately in each component of the model, allowing predictors to influence either outcome, both outcomes or neither outcome.

Our paper makes the following contributions. First, we develop a Bayesian continuous--ordinal regression model that represents academic achievement through a latent variable and models self-efficacy conditionally on latent academic achievement. This formulation allows us to directly investigate how self-efficacy varies with latent academic achievement after accounting for observed covariates. Second, we incorporate outcome-specific Bayesian variable selection, allowing predictors to influence either outcome, both outcomes or neither outcome. Third, the model structure facilitates analytical marginalisation of regression coefficients and leads to a tailored partially collapsed Gibbs sampler for posterior inference. Fourth, through an extensive simulation study, we demonstrate that the proposed sampler outperforms the more general Gaussian copula regression Bayesian variable-selection approach in \citet{alexopoulos2021bayesian}, both in terms of mixing and variable-selection performance. Finally, we apply the proposed methodology to the LSAC dataset, providing insights into the factors associated with academic achievement and self-efficacy and the relationship between these outcomes, and outlining policy-related implications arising from the findings.

The rest of the paper is organised as follows. Section \ref{sec:methodology} introduces the proposed model and Bayesian variable selection specification; related Bayesian variable selection approaches are reviewed in Section \ref{subsec:bayesian_variable_selection_review}. Section \ref{sec:inference} derives the posterior distribution and outlines the partially collapsed Gibbs sampler used for posterior inference, and contrasts our sampler with the Bayesian variable selection approach for Gaussian copula regression proposed in \citet{alexopoulos2021bayesian}. Section \ref{sec:application} presents an application to the LSAC dataset. Section \ref{sec:conclusion_future_reseach} concludes and outlines future research. Derivations of the full conditional distributions underlying the partially collapsed Gibbs sampler and an extensive simulation study comparing our approach with that of \citet{alexopoulos2021bayesian} are provided in the supplementary material. The LSAC dataset used in the application is confidential and cannot be publicly distributed. Python code implementing the proposed methodology is publicly available\footnote{Code available at \url{https://github.com/matiasq/continuous-ordinal-bvs}.}. We refer to equations and sections in the main paper as (1.1) and Section 1, respectively. Corresponding items in the supplement are denoted by (S1.1) and Section S1.

\section{Methodology} \label{sec:methodology}

\subsection{Observed responses and covariates}

Let $y_{1i}$ denote a continuous measure of self-efficacy and $y_{2i}$ an ordinal measure of academic performance with $K$ ordered categories for child $i$, $i=1,\dots,n$. Let $\mathbf{x}_i=(x_{i1},\dots,x_{ip})^\top \in \mathbb{R}^p$ denote a vector of $p$ observed covariates for child $i$. We aim to model $y_{1i}$ and $y_{2i}$ jointly conditional on $\mathbf{x}_i$, allowing for dependence between the two responses. The effects of the covariates on self-efficacy and academic performance are represented by regression coefficient vectors $\boldsymbol{\beta}_1,\boldsymbol{\beta}_2 \in \mathbb{R}^p$, respectively.

We assume throughout that the covariates $\mathbf{x}_i$ have been standardised to have mean zero and unit variance. In addition, the continuous response $y_{1i}$ is centred to have mean zero. These transformations allow the intercept term to be omitted from the regression equation for self-efficacy. Under the latent representation of the ordinal response introduced in Section~\ref{subsec:latent_representation}, a separate intercept is not identified jointly with the ordinal thresholds and is therefore omitted from the academic performance model. Using the same set of standardised covariates in both equations simplifies the model specification and notation.

\subsection{Latent-achievement model}
\label{subsec:latent_representation}
The academic-performance measure in the LSAC dataset is a five-category parental rating of the child’s progress relative to classmates. Five-category ordered rating scales are widely used in educational and psychometric measurement; they provide sufficient discrimination for respondent judgements while avoiding the assumption that respondents can reliably make very fine distinctions \citep{LozanoGarciaCuetoMuniz2008, PrestonColman2000}. Similar five-point academic rating scales are used in major longitudinal education studies such as the Early Childhood Longitudinal Study, Kindergarten Class (ECLS-K) \citep{TourangeauEtAl2009ECLSK}.

Let $y_{2i} \in {1,\dots,K=5}$ denote the observed parental assessment of academic performance for child $i$. We model $y_{2i}$  as an ordinal observation of an underlying latent construct of academic achievement. This preserves the ordered nature of the response while avoiding the assumption of equal spacing between categories inherent in a linear model. We introduce a continuous latent variable $y_{2i}^*$, interpreted as latent academic achievement as reflected in the parental assessments of child $i$, and assume that the observed academic-performance category $y_{2i}$ is determined by whether $y_{2i}^*$ falls between a set of threshold values, which are estimated from the data.

Specifically, the ordinal observed academic-performance response has probabilities
\begin{align*}
\Pr(y_{2i} = k), \quad k = 1,\dots,K.
\end{align*}
We model these probabilities by thresholding latent academic achievement $y_{2i}^*$ through
\begin{align*}
y_{2i} = k \quad \Longleftrightarrow \quad \xi_{k-1} < y_{2i}^* \le \xi_k,
\end{align*}
where
\begin{align*}
-\infty = \xi_0 < \xi_1 < \cdots < \xi_{K-1} < \xi_K = \infty.
\end{align*}
Thus, the probability of observing academic-performance category $k$ is determined by the probability that latent academic achievement lies between the corresponding threshold values, 
\begin{align*}
\Pr(y_{2i} = k)
=
\Pr(\xi_{k-1} < y_{2i}^* \le \xi_k).
\end{align*}

In a regression framework, we model latent academic achievement as
\begin{align*}
y_{2i}^* &= \mathbf{x}_i^\top \boldsymbol{\beta}_2 + e_{2i}, \\
e_{2i} &\sim \mathcal{N}(0,1),
\end{align*}
where the variance of $e_{2i}$ is fixed for identifiability. It follows that
\begin{align*}
\Pr(y_{2i} = k)
&= \Phi(\xi_k - \mathbf{x}_i^\top \boldsymbol{\beta}_2)
 - \Phi(\xi_{k-1} - \mathbf{x}_i^\top \boldsymbol{\beta}_2),
\end{align*}
where $\Phi$ denotes the cumulative distribution function of the standard normal distribution.

The latent achievement variable $y^*_{2i}$ plays a central role in our methodology because it subsequently enters the model for self-efficacy as a predictor. This yields a conditional formulation that provides an interpretable characterisation of the relationship between academic performance and self-efficacy.

\subsection{Conditional model for self-efficacy}\label{subsec:conditional_self_efficacy}

There is increasing evidence for the theorised directional relationship from perceived academic mastery to self-efficacy \citep{CrippsGraceFrancis2026EducationalAttainment}. Motivated by this finding, we model self-efficacy conditionally on latent academic achievement. Rather than representing dependence through a residual correlation, we directly relate deviations from expected self-efficacy to deviations from expected latent academic achievement. This yields an interpretable regression formulation in which the relationship between the two outcomes is represented through a regression coefficient, while allowing observed covariates to influence each outcome separately.

Specifically, from Section \ref{subsec:latent_representation}, latent academic achievement satisfies
\begin{align*}
\mathbb{E}(y_{2i}^* | \mathbf{x}_i)
=
\mathbf{x}_i^\top \boldsymbol{\beta}_2.
\end{align*}
Similarly, we define the conditional expectation of self-efficacy as
\begin{align*}
\mathbb{E}(y_{1i} | \mathbf{x}_i)
=
\mathbf{x}_i^\top \boldsymbol{\beta}_1.
\end{align*}
Consequently,
\begin{align*}
y_{2i}^* - \mathbf{x}_i^\top \boldsymbol{\beta}_2
\quad \text{and} \quad
y_{1i} - \mathbf{x}_i^\top \boldsymbol{\beta}_1
\end{align*}
represent deviations from the conditional expectations implied by the observed covariates. We thus model deviations from expected self-efficacy as a linear function of deviations from expected latent academic achievement,
\begin{align}\label{eq:conditional_model}
y_{1i}
-
\mathbf{x}_i^\top \boldsymbol{\beta}_1
=
\delta_{12}
\left(
y_{2i}^*
-
\mathbf{x}_i^\top \boldsymbol{\beta}_2
\right)
+
e_{1i},
\end{align}
where
\begin{align*}
e_{1i}
\sim
\mathcal{N}(0,\nu_{11}).
\end{align*}
The key parameter $\delta_{12}$ quantifies how deviations from expected latent parental academic assessment are associated with deviations from expected self-efficacy after accounting for the effects of the observed covariates. Positive values of $\delta_{12}$ indicate that children with higher latent parental academic assessment than expected, given their covariates, also tend to exhibit higher self-efficacy than expected. Conversely, negative values indicate that children with higher latent parental academic assessment than expected tend to exhibit lower self-efficacy than expected.

\subsection{Model}\label{subsec:model}

We now combine the latent-achievement model from Section \ref{subsec:latent_representation} and the conditional model for self-efficacy from Section \ref{subsec:conditional_self_efficacy} into a single Bayesian hierarchical model.

The observed academic-performance category is linked to latent academic achievement through the threshold representation
\begin{align*}
y_{2i} = k
\quad \Longleftrightarrow \quad
\xi_{k-1} < y_{2i}^* \le \xi_k,
\end{align*}
where
\begin{align*}
-\infty = \xi_0 < \xi_1 < \cdots < \xi_{K-1} < \xi_K = \infty.
\end{align*}
Conditional on latent academic achievement, self-efficacy follows
\begin{align*}
y_{1i}
=
\mathbf{x}_i^\top \boldsymbol{\beta}_1
+
\delta_{12}
\left(
y_{2i}^*
-
\mathbf{x}_i^\top \boldsymbol{\beta}_2
\right)
+
e_{1i},
\end{align*}
where
\begin{align*}
e_{1i}
\sim
\mathcal{N}(0,\nu_{11}).
\end{align*}
Latent academic achievement is modelled as
\begin{align}\label{eq:latent_equation}
y_{2i}^*
=
\mathbf{x}_i^\top \boldsymbol{\beta}_2
+
e_{2i},
\end{align}
where
\begin{align*}
e_{2i}
\sim
\mathcal{N}(0,1),
\end{align*}
with $e_{1i}$ and $e_{2i}$ independent. 

\begin{figure}[h]
\centering
\begin{tikzpicture}[
    node distance=1.8cm and 2.5cm,
    latent/.style={circle, draw, fill=green!18, minimum size=1.05cm, align=center},
    obs/.style={circle, draw, fill=purple!18, minimum size=1.05cm, align=center},
    param/.style={rectangle, draw, rounded corners, fill=green!18, minimum height=0.8cm, align=center},
    prior/.style={rectangle, draw, rounded corners, fill=orange!22, minimum height=0.8cm, align=center},
    plate/.style={draw, rounded corners, inner sep=0.65cm},
    arrow/.style={->, thick}
]

\node[prior] (p_beta1_delta) {$p(\boldsymbol{\beta}_1,\delta_{12}|\nu_{11})$};
\node[param, below=1.2cm of p_beta1_delta] (beta1) {$\boldsymbol{\beta}_1$};

\node[prior, right=3.1cm of p_beta1_delta] (p_beta2) {$p(\boldsymbol{\beta}_2)$};
\node[param, below=1.2cm of p_beta2] (beta2) {$\boldsymbol{\beta}_2$};

\node[param, right=2.4cm of beta2] (delta) {$\delta_{12}$};

\node[prior, right=3.2cm of delta] (p_v11) {$p(\nu_{11})$};
\node[param, below=1.2cm of p_v11] (v11) {$\nu_{11}$};

\node[prior, below=2.2cm of p_beta1_delta] (p_xi) {$p(\boldsymbol{\xi})$};
\node[param, below=1.2cm of p_xi] (xi) {$\boldsymbol{\xi}$};

\node[latent, below=1.9cm of beta2] (ystar) {$y_{2i}^*$};
\node[obs, below=1.45cm of ystar] (y2) {$y_{2i}$};
\node[obs, right=3.2cm of y2] (y1) {$y_{1i}$};

\draw[arrow] (p_beta1_delta) -- (beta1);
\draw[arrow] (p_beta1_delta) -- (delta);
\draw[arrow] (p_beta2) -- (beta2);
\draw[arrow] (p_v11) -- (v11);
\draw[arrow] (p_xi) -- (xi);

\draw[arrow] (beta2) -- (ystar);
\draw[arrow] (ystar) -- (y2);
\draw[arrow] (xi) -- (y2);

\draw[arrow] (beta1.east) to[out=-35,in=170] (y1.west);
\draw[arrow] (beta2.east) -- (y1.125);
\draw[arrow] (ystar) -- (y1);
\draw[arrow] (delta) -- (y1);
\draw[arrow] (v11.west) -- (y1.north east);

\begin{scope}[on background layer]
\node[
    plate,
    fit=(ystar)(y2)(y1),
    label={[anchor=south east, yshift=-3.5pt]south east:$i=1,\dots,n$}
] {};
\end{scope}

\end{tikzpicture}

\caption{Graphical representation of the proposed model for self-efficacy ($y_{1i}$) and academic performance ($y_{2i}$). The observed academic-performance category $y_{2i}$ for child $i$ is linked to latent academic achievement $y_{2i}^*$ through the threshold model, and latent academic achievement enters the conditional model for self-efficacy as a predictor. Shaded purple nodes denote observed variables, green nodes denote latent variables or model parameters, and orange nodes denote prior distributions. The plate indicates quantities repeated for $i=1,\dots,n$. The graph represents the model conditional on the observed covariates $\mathbf{x}_i$, which are omitted for clarity.}\label{fig:graphical_model}
\end{figure}
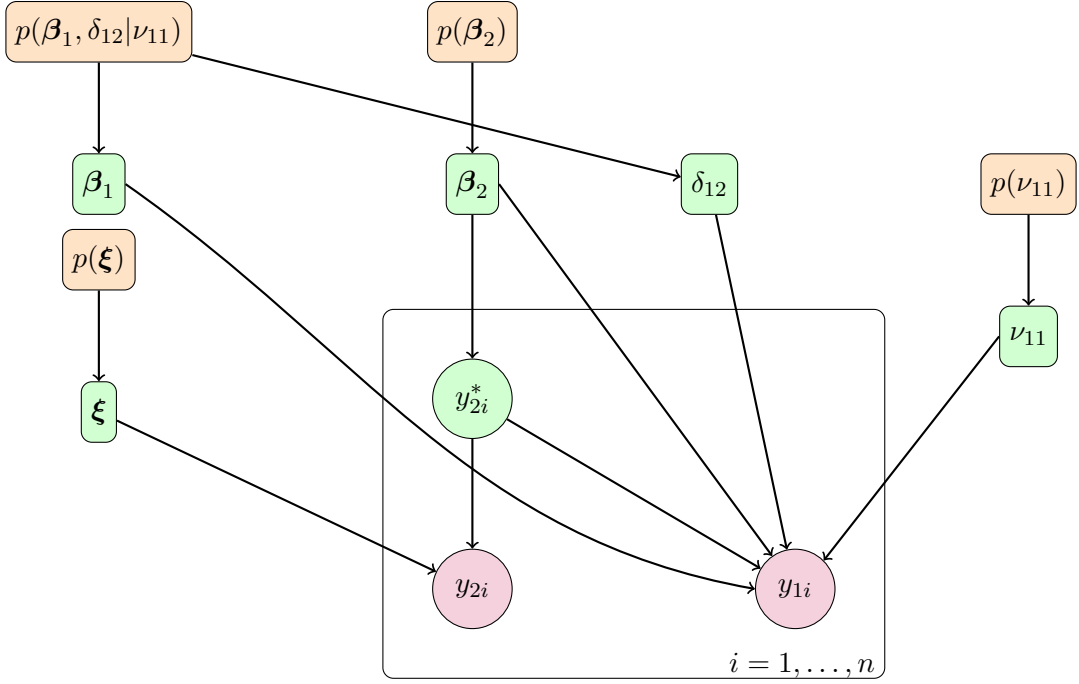

The model is completed by assigning prior distributions. Let
\begin{align*}
\boldsymbol{\xi}
=
(\xi_1,\ldots,\xi_{K-1})^\top
\end{align*}
denote the vector of threshold parameters. We assume the prior factorisation
\begin{align}\label{eq:prior_factorisation}
p(\boldsymbol{\beta}_1,\boldsymbol{\beta}_2,\delta_{12},\nu_{11},\boldsymbol{\xi})
=
p(\boldsymbol{\beta}_1,\delta_{12} | \nu_{11})
p(\boldsymbol{\beta}_2)
p(\nu_{11})
p(\boldsymbol{\xi}),
\end{align}
which reflects the hierarchical structure of the model. Together with suitably chosen prior families, this factorisation facilitates posterior computation. Figure \ref{fig:graphical_model} provides a graphical representation of the model.

\subsection{Joint model representation}\label{subsec:joint_model}

Although the proposed model is formulated via a marginal model for latent academic achievement and a conditional model for self-efficacy given latent academic achievement, it admits an equivalent joint Gaussian representation. 

Specifically, from Sections~\ref{subsec:latent_representation} and \ref{subsec:conditional_self_efficacy},
\begin{align*}
y_{2i}^*
&=
\mathbf{x}_i^\top \boldsymbol{\beta}_2
+
e_{2i},\\
y_{1i}
&=
\mathbf{x}_i^\top \boldsymbol{\beta}_1
+
\delta_{12} e_{2i}
+
e_{1i},
\end{align*}
where
\begin{align*}
e_{2i}
&\sim
\mathcal{N}(0,1),\\
e_{1i}
&\sim
\mathcal{N}(0,\nu_{11}),
\end{align*}
independently. It follows that latent academic achievement and self-efficacy jointly follow a bivariate normal distribution conditional on the observed covariates,
\begin{align}
\begin{pmatrix}
y_{1i} \\
y_{2i}^*
\end{pmatrix}
\Bigg|
\mathbf{x}_i
\sim
\mathcal{N}
\left(
\begin{pmatrix}
\mathbf{x}_i^\top \boldsymbol{\beta}_1 \\
\mathbf{x}_i^\top \boldsymbol{\beta}_2
\end{pmatrix},
\begin{pmatrix}
\nu_{11} + \delta_{12}^2 & \delta_{12} \\
\delta_{12} & 1
\end{pmatrix}
\right).
\label{eq:joint_representation}
\end{align}
Thus, the conditional formulation relating self-efficacy to latent academic achievement is statistically equivalent to a joint Gaussian model for the two outcomes. In particular, if
\begin{align*}
\begin{pmatrix}
y_{1i} \\
y_{2i}^*
\end{pmatrix}
\Bigg|
\mathbf{x}_i
\sim
\mathcal{N}
\left(
\begin{pmatrix}
\mathbf{x}_i^\top \boldsymbol{\beta}_1 \\
\mathbf{x}_i^\top \boldsymbol{\beta}_2
\end{pmatrix},
\begin{pmatrix}
\sigma_{11} & \sigma_{12} \\
\sigma_{12} & 1
\end{pmatrix}
\right),
\end{align*}
then
\begin{align*}
\sigma_{11}
&=
\nu_{11}+\delta_{12}^2,\\
\sigma_{12}
&=
\delta_{12}.
\end{align*}
Consequently, covariance-based measures of association between self-efficacy and latent academic achievement can be recovered directly from posterior samples of $(\delta_{12},\nu_{11})$. For example, the conditional correlation between self-efficacy and latent academic achievement is
\begin{align}\label{eq:conditional_corr}
\mathrm{Corr}(y_{1i},y_{2i}^* | \mathbf{x}_i)
=
\frac{\delta_{12}}
{\sqrt{\nu_{11}+\delta_{12}^2}}.
\end{align}

The joint representation in \eqref{eq:joint_representation} also reveals that the proposed model may be viewed as a special case of the more general Gaussian copula regression framework in \cite{alexopoulos2021bayesian}. The distinction of the present work is therefore not the underlying statistical model, but rather the parameterisation that facilitates a tailored computational strategy based on explicit marginalisation of regression coefficients; see Section \ref{subsec:comparison_copula} for details.

\subsection{Bayesian variable selection}\label{subsec:bayesian_variable_selection}

To incorporate variable selection, we introduce binary inclusion indicators
\begin{align*}
\boldsymbol{\gamma}_1 = (\gamma_{11},\dots,\gamma_{1p})^\top,
\qquad
\boldsymbol{\gamma}_2 = (\gamma_{21},\dots,\gamma_{2p})^\top,
\end{align*}
where
\begin{align*}
\gamma_{rj} \in \{0,1\},
\qquad
r \in \{1,2\},
\quad
j=1,\dots,p.
\end{align*}
The indicator $\gamma_{rj}$ determines whether the $j$-th covariate is included in the regression model for response $r$. Let
\[
\mathcal{A}_r=\{j:\gamma_{rj}=1\},
\qquad r=1,2,
\]
denote the corresponding active set of predictors. Given $\boldsymbol{\gamma}_r$, we define the corresponding submatrix of covariates by selecting only the columns of $\mathbf{X}$ with indices in $\mathcal{A}_r$, which we denote by $\mathbf{X}_{\mathcal{A}_r}$. Similarly, $\boldsymbol{\beta}_{r,\mathcal{A}_r}$ denotes the subvector of regression coefficients corresponding to the active predictors in $\mathcal{A}_r$. Coefficients corresponding to excluded covariates are fixed at zero, so that
\begin{align*}
\beta_{rj}=0
\qquad
\text{whenever}
\qquad
\gamma_{rj}=0.
\end{align*}
The indicators in $\boldsymbol{\gamma}_2$ determine which observed covariates are associated with latent academic achievement, whereas the indicators in $\boldsymbol{\gamma}_1$ determine which observed covariates are associated with self-efficacy after accounting for latent academic achievement. Consequently, a covariate may influence academic achievement, self-efficacy, both outcomes, or neither outcome.

For the latent-achievement component of the model, variable selection is performed on the regression coefficients $\boldsymbol{\beta}_2$. Conditional on $\boldsymbol{\gamma}_2$, the regression coefficients $\boldsymbol{\beta}_{2,\mathcal{A}_2}$ are assigned a Zellner--$g$ prior \citep{zellner1986gprior},
\begin{align*}
\boldsymbol{\beta}_{2,\mathcal{A}_2}
\sim
\mathcal{N}
\left(
\mathbf{0},
\;
g_2
\left(
\mathbf{X}_{\mathcal{A}_2}^\top
\mathbf{X}_{\mathcal{A}_2}
\right)^{-1}
\right),
\qquad
g_2 > 0.
\end{align*}

For the self-efficacy component, variable selection is performed on the regression coefficients $\boldsymbol{\beta}_1$, while the association parameter $\delta_{12}$ is always included in the model. As reflected in the prior factorisation (\ref{eq:prior_factorisation}), the active coefficients $\boldsymbol{\beta}_{1,\mathcal{A}_1}$ and the association parameter $\delta_{12}$ are assigned a joint prior distribution conditional on $\nu_{11}$. To accommodate this structure, we augment the design matrix with the term
\begin{align*}
\mathbf{y}_2^*
-
\mathbf{X}_{\mathcal{A}_2}
\boldsymbol{\beta}_{2,\mathcal{A}_2}
\end{align*}
and define the corresponding augmented design matrix
\begin{align*}
\widetilde{\mathbf{X}}_{\mathcal{A}_1}
=
\left[
\mathbf{X}_{\mathcal{A}_1}
\;\;
\mathbf{y}_2^*
-
\mathbf{X}_{\mathcal{A}_2}
\boldsymbol{\beta}_{2,\mathcal{A}_2}
\right].
\end{align*}
Conditional on $\boldsymbol{\gamma}_1$ and $\nu_{11}$, the active regression coefficients $\boldsymbol{\beta}_{1,\mathcal{A}_1}$ and the association parameter $\delta_{12}$ are assigned the joint Zellner--$g$ prior
\begin{align*}
\widetilde{\boldsymbol{\beta}}
=
\left(
\boldsymbol{\beta}_{1,\mathcal{A}_1},
\delta_{12}
\right)^\top
\sim
\mathcal{N}
\left(
\mathbf{0},
\;
g_1 \nu_{11}
\left(
\widetilde{\mathbf{X}}_{\mathcal{A}_1}^\top
\widetilde{\mathbf{X}}_{\mathcal{A}_1}
\right)^{-1}
\right),
\qquad
g_1 > 0.
\end{align*}

Finally, we place independent hierarchical Bernoulli priors on the inclusion indicators:
\begin{align*}
\gamma_{rj}
\sim
\mathrm{Bernoulli}(\pi_r),
\qquad
r \in \{1,2\},
\quad
j=1,\dots,p,
\end{align*}
with
\begin{align*}
\pi_r
\sim
\mathrm{Beta}(a_{\pi,r}, b_{\pi,r}),
\qquad
r \in \{1,2\}.
\end{align*}
The hierarchical beta--Bernoulli specification allows the degree of sparsity to be learned from the data separately for the academic-achievement and self-efficacy components of the model. Posterior inference therefore identifies covariates that are important for explaining latent academic achievement, self-efficacy, or both outcomes simultaneously.

\subsection{Related Bayesian variable selection approaches}
\label{subsec:bayesian_variable_selection_review} 

We restrict attention to Bayesian variable-selection methodologies based on exact spike-and-slab priors that induce a posterior distribution over the discrete model space. Such formulations provide posterior inclusion probabilities with a direct model-averaging interpretation and naturally quantify uncertainty regarding predictor inclusion. Early examples include linear regression \citep{mitchell1988bayesian} and nonparametric regression \citep{smith1996nonparameteric}; see \cite{ohara2009review} for a review.

Bayesian variable selection has also been studied in models where latent-variable representations facilitate posterior computation. Examples include ordinal regression in \citet{kwon2007biomarkers} through the latent Gaussian formulation in \citet{albert1993bayesian}, and logistic regression in \citet{bogoni2025bayesian} through P\'{o}lya--Gamma augmentation \citep{polson2013bayesian}. More generally, \citet{alexopoulos2021bayesian} develop Bayesian variable selection for Gaussian copula regression models, providing a flexible framework for mixed-response settings through a latent Gaussian representation. The methodology proposed in this paper similarly exploits the latent Gaussian representation in \citet{albert1993bayesian}, but is tailored to the continuous--ordinal setting arising in the LSAC application. This specialisation facilitates analytical marginalisation and the development of an efficient partially collapsed Gibbs sampler; see Section \ref{subsec:comparison_copula} for a contrast to \citet{alexopoulos2021bayesian}. The empirical performance of the two approaches is compared in the simulation study reported in Section \ref{supp:simulation_study}.

\section{Inference}\label{sec:inference}

\subsection{Posterior distribution}\label{subsec:posterior_distribution}

We now derive the posterior distribution corresponding to the model introduced in Section~\ref{subsec:model} and represented graphically in Figure~\ref{fig:graphical_model}.

Let
\begin{align*}
\mathbf{y}_1 &= (y_{11},\dots,y_{1n})^\top, \\
\mathbf{y}_2 &= (y_{21},\dots,y_{2n})^\top, \\
\mathbf{y}_2^* &= (y_{21}^*,\dots,y_{2n}^*)^\top,
\end{align*}
where $\mathbf{y}_1$ and $\mathbf{y}_2$ are the observed responses and $\mathbf{y}_2^*$ denotes the latent academic-achievement vector. By Bayes' theorem, the posterior distribution is
\begin{align}
&p(\mathbf{y}_2^*, \boldsymbol{\beta}_1, \boldsymbol{\beta}_2,
\boldsymbol{\xi}, \delta_{12}, \nu_{11}
|
\mathbf{y}_1, \mathbf{y}_2)
\nonumber\\
&\propto
p(\mathbf{y}_1,\mathbf{y}_2
|
\mathbf{y}_2^*,
\boldsymbol{\beta}_1,
\boldsymbol{\beta}_2,
\boldsymbol{\xi},
\delta_{12},
\nu_{11})
\;
p(\mathbf{y}_2^*,
\boldsymbol{\beta}_1,
\boldsymbol{\beta}_2,
\boldsymbol{\xi},
\delta_{12},
\nu_{11}).
\label{eq:joint_posterior}
\end{align}
According to the conditional independence structure encoded in Figure~\ref{fig:graphical_model}, the likelihood term in \eqref{eq:joint_posterior} factorises as
\begin{align*}
&p(\mathbf{y}_1,\mathbf{y}_2
|
\mathbf{y}_2^*,
\boldsymbol{\beta}_1,
\boldsymbol{\beta}_2,
\boldsymbol{\xi},
\delta_{12},
\nu_{11})
\\
&=
p(\mathbf{y}_2
|
\mathbf{y}_2^*,
\boldsymbol{\xi})
\;
p(\mathbf{y}_1
|
\mathbf{y}_2^*,
\boldsymbol{\beta}_1,
\boldsymbol{\beta}_2,
\delta_{12},
\nu_{11}).
\end{align*}
Since $\mathbf{y}_2$ is deterministically determined by $\mathbf{y}_2^*$ and $\boldsymbol{\xi}$ through the threshold representation, the first term reduces to an indicator function enforcing the threshold constraints. Therefore,
\begin{align*}
&p(\mathbf{y}_1,\mathbf{y}_2
|
\mathbf{y}_2^*,
\boldsymbol{\beta}_1,
\boldsymbol{\beta}_2,
\boldsymbol{\xi},
\delta_{12},
\nu_{11})
\\
&\propto
p(\mathbf{y}_1
|
\mathbf{y}_2^*,
\boldsymbol{\beta}_1,
\boldsymbol{\beta}_2,
\delta_{12},
\nu_{11}).
\end{align*}
Similarly, the graphical model in Figure \ref{fig:graphical_model}, together with the prior factorisation in \eqref{eq:prior_factorisation}, implies that the prior term in \eqref{eq:joint_posterior} factorises as
\begin{align*}
&p(\mathbf{y}_2^*,
\boldsymbol{\beta}_1,
\boldsymbol{\beta}_2,
\boldsymbol{\xi},
\delta_{12},
\nu_{11})
\\
&=
p(\boldsymbol{\beta}_1,\delta_{12}|\nu_{11})
\;
p(\nu_{11})
\;
p(\boldsymbol{\beta}_2)
\;
p(\mathbf{y}_2^*|\boldsymbol{\beta}_2)
\;
p(\boldsymbol{\xi}).
\end{align*}

Substituting the likelihood and prior factorisations into \eqref{eq:joint_posterior} yields the posterior distribution of all unknown quantities. Because this posterior distribution does not admit a closed-form expression, inference is performed using Markov chain Monte Carlo methods. In the next subsection, we develop a tailored partially collapsed Gibbs sampler that combines Gibbs updates with analytical marginalisation of selected model parameters. This reduces the dimension of the sampling space, improves mixing, and facilitates Bayesian variable selection.

\subsection{A partially collapsed Gibbs sampler for Bayesian variable selection} \label{subsec:partially_collapsed_Gibbs}

To perform posterior inference under the Bayesian variable selection framework, we augment the posterior distribution in \eqref{eq:joint_posterior} with the inclusion indicators $\boldsymbol{\gamma}_1$ and $\boldsymbol{\gamma}_2$ introduced in Section \ref{subsec:bayesian_variable_selection}. To facilitate exploration of the indicator space, we analytically integrate out the corresponding regression coefficients when updating the inclusion indicators. In particular, $(\boldsymbol{\beta}_1, \delta_{12})$ are integrated out when sampling $\boldsymbol{\gamma}_1$, and $\boldsymbol{\beta}_2$ is integrated out when sampling $\boldsymbol{\gamma}_2$. This results in collapsed updates based on marginal full conditional distributions, which typically exhibit improved mixing behaviour compared to their non-collapsed counterparts \citep{van2008partially}.

However, this strategy leads to a partially collapsed Gibbs sampler, in which different blocks are updated using full conditional distributions derived from different marginalisations of the joint posterior. In such settings, care must be taken with the ordering of updates to ensure that the Markov chain leaves the desired posterior distribution invariant \citep{van2008partially}. In particular, one must avoid conditioning on parameters that have been integrated out in previous steps unless they have subsequently been updated. To ensure validity, we adopt an ordering in which each inclusion-indicator update is immediately followed by an update of the parameters that were integrated out in that step, before these parameters appear again in subsequent conditioning sets. This ensures that each update is consistent with the appropriate marginal or conditional distribution.

The resulting partially collapsed Gibbs sampler is summarised in Algorithm \ref{alg:partially_collapsed_Gibbs}, with an ordering of updates that preserves the validity of the Markov chain. The derivation of each full conditional distribution is provided in Section \ref{app:derivation_Gibbs}. The inclusion indicators $\boldsymbol{\gamma}_1$ and $\boldsymbol{\gamma}_2$ are updated component-wise using Gibbs updates within the partially collapsed Gibbs sampler. In Algorithm \ref{alg:partially_collapsed_Gibbs}, we explicitly display the conditioning sets for each update. Parameters are omitted only when they have been analytically integrated out and are otherwise retained regardless of the conditional independence implied by Figure \ref{fig:graphical_model}. This notation makes the marginalisations underlying the partially collapsed sampler transparent.

\begin{algorithm}
\caption{Partially collapsed Gibbs sampler to sample $M$ draws from the posterior distribution of the bivariate continuous--ordinal regression model with variable selection. Full conditional distributions are given in Section \ref{app:derivation_Gibbs}.}
\label{alg:partially_collapsed_Gibbs}
\begin{algorithmic}[1]
\renewcommand{\baselinestretch}{1.2}\selectfont

\State Initialise $\boldsymbol{\beta}_2$, $\boldsymbol{\gamma}_1$, $\boldsymbol{\gamma}_2$, $\mathbf{y}_2^*$, $\nu_{11}$, $\boldsymbol{\xi}$, $\pi_1$ and $\pi_2$.

\For{$m = 1,\dots,M$}

    \State Sample $\boldsymbol{\gamma}_1$ from
    \begin{align*}
    p(\boldsymbol{\gamma}_1 | \boldsymbol{\beta}_2, \boldsymbol{\gamma}_2, \mathbf{y}_2^*, \nu_{11}, \boldsymbol{\xi}, \mathbf{y}_1, \mathbf{y}_2, \pi_1),
    \end{align*}
    analytically integrating out $(\boldsymbol{\beta}_1, \delta_{12})$.

    \State Sample $(\boldsymbol{\beta}_1, \delta_{12})$ from
    \begin{align*}
    p(\boldsymbol{\beta}_1, \delta_{12} | \boldsymbol{\beta}_2, \boldsymbol{\gamma}_1, \boldsymbol{\gamma}_2, \mathbf{y}_2^*, \nu_{11}, \boldsymbol{\xi}, \mathbf{y}_1, \mathbf{y}_2).
    \end{align*}

    \State Sample $\pi_1$ from
    \begin{align*}
    p(\pi_1 | \boldsymbol{\beta}_1, \boldsymbol{\beta}_2, \delta_{12}, \boldsymbol{\gamma}_1, \boldsymbol{\gamma}_2, \mathbf{y}_2^*, \nu_{11}, \boldsymbol{\xi}, \mathbf{y}_1, \mathbf{y}_2).
    \end{align*}

    \State Sample $\boldsymbol{\gamma}_2$ from
    \begin{align*}
    p(\boldsymbol{\gamma}_2 | \boldsymbol{\beta}_1, \delta_{12}, \boldsymbol{\gamma}_1, \mathbf{y}_2^*, \nu_{11}, \boldsymbol{\xi}, \mathbf{y}_1, \mathbf{y}_2, \pi_2),
    \end{align*}
    analytically integrating out $\boldsymbol{\beta}_2$.

    \State Sample $\boldsymbol{\beta}_2$ from
    \begin{align*}
    p(\boldsymbol{\beta}_2 | \boldsymbol{\beta}_1, \delta_{12}, \boldsymbol{\gamma}_1, \boldsymbol{\gamma}_2, \mathbf{y}_2^*, \nu_{11}, \boldsymbol{\xi}, \mathbf{y}_1, \mathbf{y}_2).
    \end{align*}

    \State Sample $\pi_2$ from
    \begin{align*}
    p(\pi_2 | \boldsymbol{\beta}_1, \boldsymbol{\beta}_2, \delta_{12}, \boldsymbol{\gamma}_1, \boldsymbol{\gamma}_2, \mathbf{y}_2^*, \nu_{11}, \boldsymbol{\xi}, \mathbf{y}_1, \mathbf{y}_2).
    \end{align*}

    \State Sample $\mathbf{y}_2^*$ from
    \begin{align*}
    p(\mathbf{y}_2^* | \boldsymbol{\beta}_1, \boldsymbol{\beta}_2, \delta_{12}, \boldsymbol{\gamma}_1, \boldsymbol{\gamma}_2, \nu_{11}, \boldsymbol{\xi}, \mathbf{y}_1, \mathbf{y}_2).
    \end{align*}

    \State Sample $\nu_{11}$ from
    \begin{align*}
    p(\nu_{11} | \boldsymbol{\beta}_1, \boldsymbol{\beta}_2, \delta_{12}, \boldsymbol{\gamma}_1, \boldsymbol{\gamma}_2, \mathbf{y}_2^*, \boldsymbol{\xi}, \mathbf{y}_1, \mathbf{y}_2).
    \end{align*}

    \State Sample $\boldsymbol{\xi}$ from
    \begin{align*}
    p(\boldsymbol{\xi} | \boldsymbol{\beta}_1, \boldsymbol{\beta}_2, \delta_{12}, \boldsymbol{\gamma}_1, \boldsymbol{\gamma}_2, \mathbf{y}_2^*, \nu_{11}, \mathbf{y}_1, \mathbf{y}_2).
    \end{align*}

\EndFor

\end{algorithmic}
\end{algorithm}

\subsection{Comparison with \citet{alexopoulos2021bayesian}}\label{subsec:comparison_copula}

As discussed in Section \ref{subsec:joint_model}, the corresponding joint formulation of our model is a special case of a Gaussian copula regression model. A general Bayesian variable selection methodology for multivariate Gaussian copula regression models is proposed in \citet{alexopoulos2021bayesian}. In this subsection, we contrast our parameterisation and inferential strategy with their approach.

Although the joint and conditional formulations belong to the same latent Gaussian family, they differ substantially in how dependence between responses is represented. In the Gaussian copula framework of \citet{alexopoulos2021bayesian}, dependence is parameterised through a correlation matrix $\mathbf{R}$, or equivalently a precision matrix, with sparsity structures imposed through graph-based priors \citep{talhouk2012efficient}. Thus, even in the bivariate setting considered here, dependence is parameterised through covariance or precision parameters. These quantities characterise the association between the continuous response and the latent Gaussian representation of the ordinal response. In contrast, the conditional formulation adopted in this paper parameterises dependence through the regression coefficient $\delta_{12}$. As discussed in Section \ref{subsec:joint_model}, this representation is equivalent to the corresponding bivariate Gaussian copula formulation, but admits a direct interpretation as the association between latent academic achievement and self-efficacy after accounting for the observed covariates.

While the two parameterisations are statistically equivalent, a more consequential difference arises from the resulting computational strategy. The conditional formulation facilitates analytical marginalisation of the regression coefficients when updating the inclusion indicators $\boldsymbol{\gamma}_1$ and $\boldsymbol{\gamma}_2$, leading to the partially collapsed sampling scheme described in Algorithm \ref{alg:partially_collapsed_Gibbs}. In the implementation considered here, the inclusion indicators are updated through Gibbs steps based on the resulting marginal conditional distributions. In contrast, \citet{alexopoulos2021bayesian} employ a Metropolis--Hastings step for joint updates of $(\boldsymbol{\beta}_k,\boldsymbol{\gamma}_k)$. The resulting acceptance probability exhibits an implicit marginalisation of the regression coefficients \citep{held2006bayesian}, but the proposal distribution for the regression coefficients is constructed by replacing the observed-data likelihood with that of the corresponding latent Gaussian variable. This yields a quadratic form in $\boldsymbol{\beta}_k$ and therefore a closed-form proposal density. Although the Metropolis--Hastings correction preserves the desired target distribution, the efficiency of the proposal depends on how accurately the latent Gaussian representation reflects the information contained in the observed data.

In summary, both approaches arise from the same latent Gaussian framework. The conditional formulation adopted in this paper is motivated by the specific continuous--ordinal structure arising in our application. By specialising to this setting, we obtain a direct interpretation of the dependence parameter and an efficient sampling strategy that exploits this structure through analytical marginalisation. In Section \ref{supp:simulation_study}, we benchmark the two approaches using an extensive simulation study based on settings calibrated to the LSAC dataset. Across 100 simulated datasets and a range of signal-to-noise ratios, the proposed sampler exhibits improved mixing and more accurate variable selection for a fixed number of Markov chain iterations. We stress, however, that the Gaussian copula regression approach in \citet{alexopoulos2021bayesian} is considerably more general and applies to a broad range of mixed-response settings, whereas the methodology proposed in this paper is tailored to the continuous-ordinal setting relevant for our application.

\section{Academic achievement and self-efficacy: Evidence from Australian children}\label{sec:application}

\subsection{The longitudinal study of Australian children}

The Longitudinal Study of Australian Children (LSAC) is a nationally representative longitudinal study of Australian children and their families \citep{sanson2002lsac}. We analyse data from the K cohort at Wave~5, when participants were approximately 12--13 years old \citep{LSAC_Wave5_2012}.

Wave~5 contains detailed information on children's academic performance, self-efficacy, family background, health, school environment and behavioural characteristics. This corresponds to early adolescence and the first year of secondary school (Year~7), when self-efficacy, educational outcomes and students' sense of belonging at school are of particular interest to educational researchers and policy makers \citep{eccles1993development,CrippsGraceFrancis2026EducationalAttainment}. Wave~5 is the only LSAC wave containing the variable ``psychological sense of school membership'' for Year~7 students; throughout the paper, we refer to this construct simply as ``sense of belonging at school''.

\subsection{Settings and prior specification}

The analysed LSAC wave consists of $n=2{,}921$ observations. Self-efficacy and ordinal academic performance are modelled as the two responses using $p=20$ candidate predictors; see Table~\ref{tab:real-data-all-pips} for the predictor names.

Posterior inference is based on the proposed partially collapsed Gibbs sampler with $4{,}000$ burn-in iterations followed by $40{,}000$ post-burn-in iterations and no thinning.

The priors are described in Section~\ref{app:derivation_Gibbs}. Hyperparameters are fixed at
\[
g_1=n,\quad
g_2=n,\quad
a_{\pi,1}=b_{\pi,1}=a_{\pi,2}=b_{\pi,2}=1,\quad
a_{\nu}=b_{\nu}=1,
\]
where $g_1=g_2=n$ corresponds to the unit-information prior \citep{kass1995reference}. Setting $a_{\pi,r}=b_{\pi,r}=1$ induces uniform priors on the inclusion probabilities $\pi_r$, while $a_{\nu}=b_{\nu}=1$ gives a weakly informative inverse-gamma prior with a heavy right tail for $\nu_{11}$.

\subsection{Validation via posterior predictive checks}\label{subsec:validation_posterior_predictive}

We assess the adequacy of the fitted model using three posterior predictive checks that evaluate its ability to reproduce the marginal distributions of academic achievement and self-efficacy, as well as the relationship between the two outcomes.

\begin{figure}[h]
\centering

\begin{subfigure}[t]{0.48\textwidth}
    \centering
    \includegraphics[width=\textwidth]{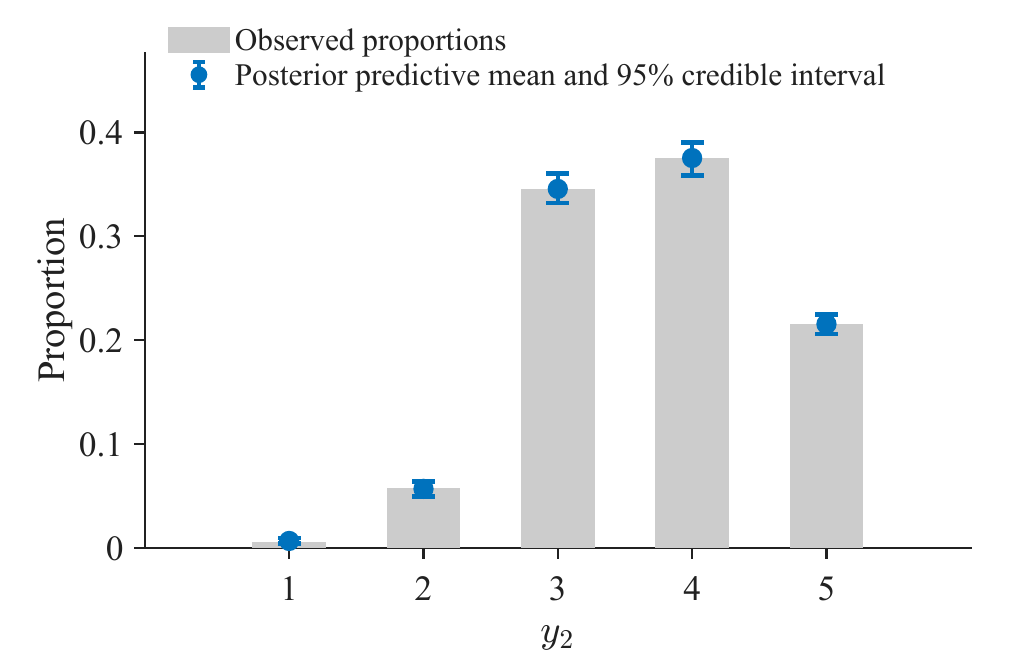}
    \caption{Academic achievement.}
    \label{fig:validation_predictive_checks_a}
\end{subfigure}
\hfill
\begin{subfigure}[t]{0.48\textwidth}
    \centering
    \includegraphics[width=\textwidth]{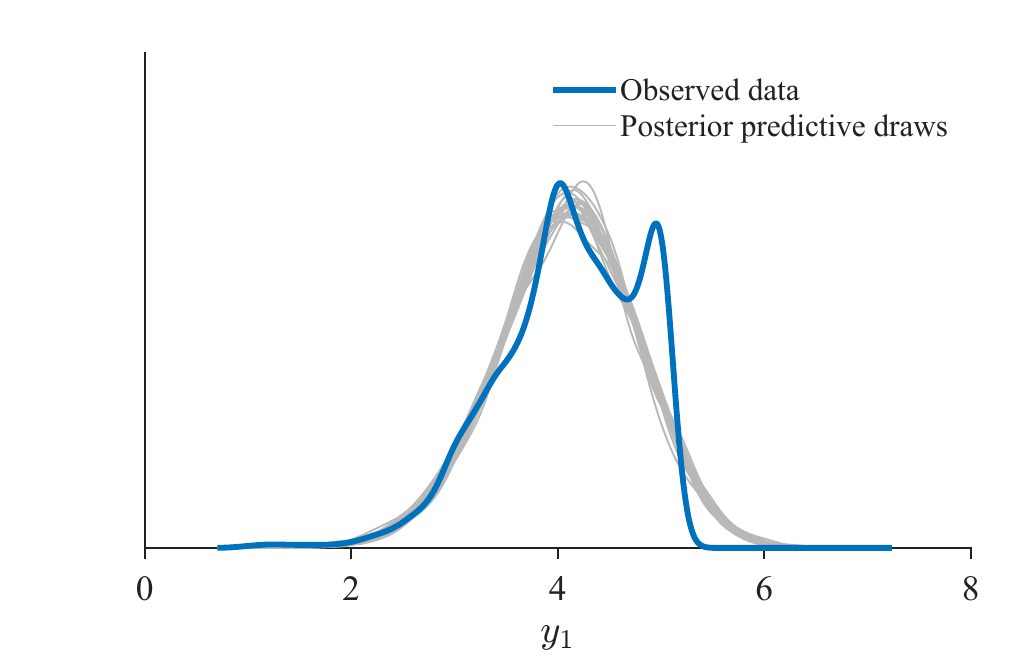}
    \caption{Self-efficacy.}
    \label{fig:validation_predictive_checks_b}
\end{subfigure}

\vspace{0.8em}

\begin{subfigure}[t]{0.48\textwidth}
    \centering
    \includegraphics[width=\textwidth]{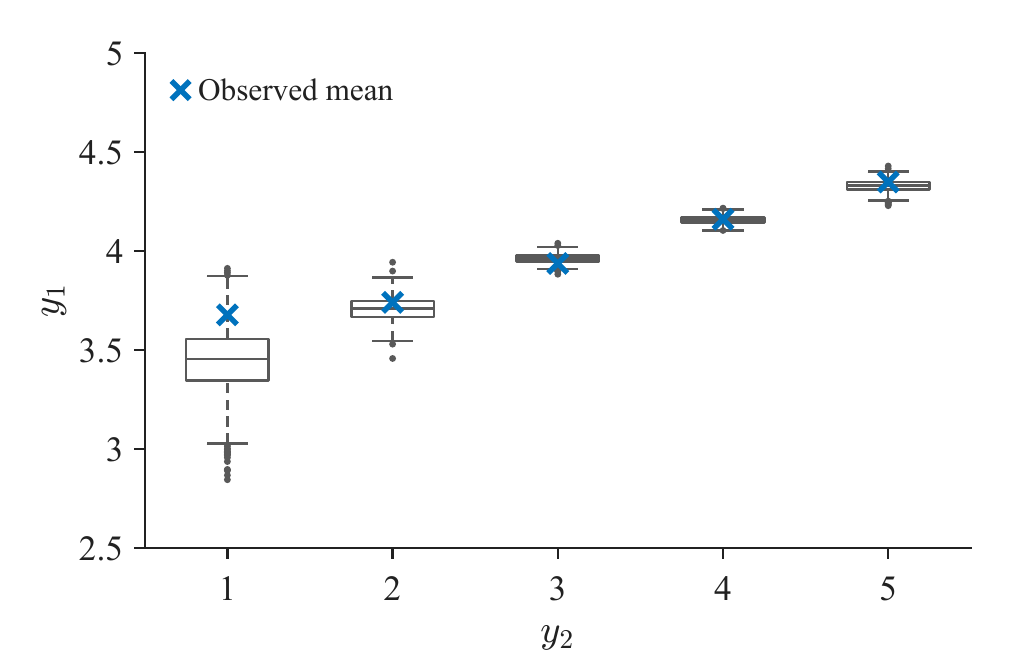}
    \caption{Self-efficacy by academic achievement.}
    \label{fig:validation_predictive_checks_c}
\end{subfigure}

\caption{Posterior predictive checks for the fitted model. Panel~(a) shows posterior predictive distributions for the proportions in the five academic achievement categories ($1{,}000$ posterior draws). Panel~(b) compares the kernel density estimate of the observed self-efficacy data with kernel density estimates from $20$ datasets generated from posterior draws (shown for visual clarity). Panel~(c) shows boxplots summarising the posterior predictive distributions for the mean self-efficacy within each academic achievement category ($1{,}000$ posterior draws).}\label{fig:validation_predictive_checks}
\end{figure}

Figure~\ref{fig:validation_predictive_checks}(\subref*{fig:validation_predictive_checks_a}) compares the observed proportions for each academic achievement category with the corresponding posterior predictive distributions. The observed proportions closely match the posterior predictive means, with the 95\% credible intervals indicating close agreement across all five categories. The fitted model therefore adequately reproduces the marginal distribution of academic achievement.

Figure~\ref{fig:validation_predictive_checks}(\subref*{fig:validation_predictive_checks_b}) compares the kernel density estimate of the observed self-efficacy data with kernel density estimates of datasets generated from posterior draws. The model reproduces the overall location and spread of the distribution reasonably well but does not capture the pronounced second mode corresponding to children with particularly high self-efficacy, most likely due to the Gaussian assumption for the continuous response. A finite mixture of Gaussian distributions could provide a more flexible model.

Figure~\ref{fig:validation_predictive_checks}(\subref*{fig:validation_predictive_checks_c}) compares the observed mean self-efficacy within each academic achievement category with the corresponding posterior predictive distributions. The observed means closely match the posterior predictive distributions across the five categories, apart from a slight deviation in the first category due to its relatively small number of observations. Overall, the fitted model adequately captures the relationship between academic achievement and self-efficacy.

Taken together, these posterior predictive checks suggest that the proposed model provides an adequate description of the LSAC data analysed in this study.

\subsection{Variable selection results}
Our analysis first examines which predictors are associated with self-efficacy and latent academic achievement. Table~\ref{tab:real-data-all-pips} summarises the posterior inclusion probabilities and estimated effects for all predictors, allowing us to identify the variables most strongly supported by the data; see the table caption for details of the reported quantities.

\begin{table}
    \centering
\caption{Posterior inclusion probabilities (PIP) and effect estimates for each predictor of self-efficacy and latent academic achievement in the LSAC dataset. Effect estimates are posterior means conditional on inclusion, $\mathbb{E}(\beta_{rj}| \gamma_{rj}=1,\mathbf{y})$, $r=1,2$, reported on the original covariate scale, with 95\% credible intervals shown below in brackets. Posterior inclusion probabilities exceeding 0.5 are shown in bold. Values less than 0.001 in absolute value are reported as 0.000. SDQ denotes the strengths and difficulties questionnaire \citep{goodman1997strengths}. ADHD/ADD denotes attention-deficit/hyperactivity disorder or attention deficit disorder.}
\label{tab:real-data-all-pips}
\footnotesize
\begin{adjustbox}{center,max width=\textwidth}
\begin{tabular}{lcccc}
\toprule
Predictor
& \multicolumn{2}{c}{Self-efficacy}
& \multicolumn{2}{c}{Latent academic achievement} \\
\cmidrule(lr){2-3}\cmidrule(lr){4-5}
& PIP & $\hat{\beta}_1$
& PIP & $\hat{\beta}_2$ \\
\midrule

SDQ conduct problems
& $\mathbf{1.000}$ & \begin{tabular}[t]{@{}c@{}}$-0.048$\\[-1mm]{\scriptsize $[-0.064,-0.032]$}\end{tabular}
& $\mathbf{1.000}$ & \begin{tabular}[t]{@{}c@{}}$-0.115$\\[-1mm]{\scriptsize $[-0.145,-0.084]$}\end{tabular} \\

\begin{tabular}[t]{@{}l@{}}Psychological sense of\\school membership\end{tabular}
& $\mathbf{1.000}$ & \begin{tabular}[t]{@{}c@{}}$0.044$\\[-1mm]{\scriptsize $[0.041,0.048]$}\end{tabular}
& $\mathbf{1.000}$ & \begin{tabular}[t]{@{}c@{}}$0.018$\\[-1mm]{\scriptsize $[0.012,0.024]$}\end{tabular} \\

Frequent school absence
& $0.006$ & \begin{tabular}[t]{@{}c@{}}$-0.020$\\[-1mm]{\scriptsize $[-0.093,0.058]$}\end{tabular}
& $\mathbf{0.987}$ & \begin{tabular}[t]{@{}c@{}}$-0.328$\\[-1mm]{\scriptsize $[-0.489,-0.169]$}\end{tabular} \\

Depression
& $0.057$ & \begin{tabular}[t]{@{}c@{}}$-0.056$\\[-1mm]{\scriptsize $[-0.108,-0.005]$}\end{tabular}
& $0.033$ & \begin{tabular}[t]{@{}c@{}}$0.024$\\[-1mm]{\scriptsize $[-0.075,0.127]$}\end{tabular} \\

Parent communication
& $0.032$ & \begin{tabular}[t]{@{}c@{}}$0.040$\\[-1mm]{\scriptsize $[-0.003,0.085]$}\end{tabular}
& $0.163$ & \begin{tabular}[t]{@{}c@{}}$0.087$\\[-1mm]{\scriptsize $[-0.001,0.173]$}\end{tabular} \\

Teacher communication
& $0.006$ & \begin{tabular}[t]{@{}c@{}}$-0.001$\\[-1mm]{\scriptsize $[-0.028,0.027]$}\end{tabular}
& $\mathbf{0.998}$ & \begin{tabular}[t]{@{}c@{}}$0.120$\\[-1mm]{\scriptsize $[0.067,0.172]$}\end{tabular} \\

General health
& $0.009$ & \begin{tabular}[t]{@{}c@{}}$0.012$\\[-1mm]{\scriptsize $[-0.019,0.041]$}\end{tabular}
& $\mathbf{1.000}$ & \begin{tabular}[t]{@{}c@{}}$0.175$\\[-1mm]{\scriptsize $[0.119,0.232]$}\end{tabular} \\

Screen time
& $0.011$ & \begin{tabular}[t]{@{}c@{}}$-0.000$\\[-1mm]{\scriptsize $[-0.000,0.000]$}\end{tabular}
& $\mathbf{0.994}$ & \begin{tabular}[t]{@{}c@{}}$-0.000$\\[-1mm]{\scriptsize $[-0.000,-0.000]$}\end{tabular} \\

One good friend or more
& $\mathbf{0.988}$ & \begin{tabular}[t]{@{}c@{}}$0.133$\\[-1mm]{\scriptsize $[0.074,0.191]$}\end{tabular}
& $0.354$ & \begin{tabular}[t]{@{}c@{}}$0.142$\\[-1mm]{\scriptsize $[0.030,0.255]$}\end{tabular} \\

Victim of bullying
& $0.247$ & \begin{tabular}[t]{@{}c@{}}$0.079$\\[-1mm]{\scriptsize $[0.024,0.133]$}\end{tabular}
& $0.085$ & \begin{tabular}[t]{@{}c@{}}$0.082$\\[-1mm]{\scriptsize $[-0.025,0.185]$}\end{tabular} \\

Any sleep problem
& $0.007$ & \begin{tabular}[t]{@{}c@{}}$0.010$\\[-1mm]{\scriptsize $[-0.037,0.059]$}\end{tabular}
& $0.062$ & \begin{tabular}[t]{@{}c@{}}$-0.061$\\[-1mm]{\scriptsize $[-0.152,0.030]$}\end{tabular} \\

Family cohesion
& $0.006$ & \begin{tabular}[t]{@{}c@{}}$-0.001$\\[-1mm]{\scriptsize $[-0.024,0.023]$}\end{tabular}
& $0.151$ & \begin{tabular}[t]{@{}c@{}}$0.043$\\[-1mm]{\scriptsize $[0.001,0.087]$}\end{tabular} \\

Extracurricular activities
& $\mathbf{0.590}$ & \begin{tabular}[t]{@{}c@{}}$0.128$\\[-1mm]{\scriptsize $[0.054,0.203]$}\end{tabular}
& $0.419$ & \begin{tabular}[t]{@{}c@{}}$0.197$\\[-1mm]{\scriptsize $[0.046,0.346]$}\end{tabular} \\

\begin{tabular}[t]{@{}l@{}}Neighbourhood social\\capital\end{tabular}
& $0.006$ & \begin{tabular}[t]{@{}c@{}}$-0.007$\\[-1mm]{\scriptsize $[-0.033,0.021]$}\end{tabular}
& $0.036$ & \begin{tabular}[t]{@{}c@{}}$0.020$\\[-1mm]{\scriptsize $[-0.036,0.079]$}\end{tabular} \\

\begin{tabular}[t]{@{}l@{}}Ongoing medical\\condition\end{tabular}
& $0.151$ & \begin{tabular}[t]{@{}c@{}}$0.055$\\[-1mm]{\scriptsize $[0.013,0.096]$}\end{tabular}
& $0.361$ & \begin{tabular}[t]{@{}c@{}}$0.108$\\[-1mm]{\scriptsize $[0.023,0.195]$}\end{tabular} \\

Student absence
& $0.007$ & \begin{tabular}[t]{@{}c@{}}$0.002$\\[-1mm]{\scriptsize $[-0.010,0.012]$}\end{tabular}
& $\mathbf{0.915}$ & \begin{tabular}[t]{@{}c@{}}$-0.035$\\[-1mm]{\scriptsize $[-0.055,-0.015]$}\end{tabular} \\

ADHD/ADD
& $0.008$ & \begin{tabular}[t]{@{}c@{}}$0.051$\\[-1mm]{\scriptsize $[-0.073,0.160]$}\end{tabular}
& $\mathbf{0.984}$ & \begin{tabular}[t]{@{}c@{}}$-0.478$\\[-1mm]{\scriptsize $[-0.718,-0.238]$}\end{tabular} \\

Autism
& $0.014$ & \begin{tabular}[t]{@{}c@{}}$0.089$\\[-1mm]{\scriptsize $[-0.054,0.216]$}\end{tabular}
& $\mathbf{0.608}$ & \begin{tabular}[t]{@{}c@{}}$-0.398$\\[-1mm]{\scriptsize $[-0.674,-0.129]$}\end{tabular} \\

Parent education
& $0.008$ & \begin{tabular}[t]{@{}c@{}}$0.003$\\[-1mm]{\scriptsize $[-0.004,0.011]$}\end{tabular}
& $\mathbf{1.000}$ & \begin{tabular}[t]{@{}c@{}}$0.055$\\[-1mm]{\scriptsize $[0.038,0.072]$}\end{tabular} \\

Family income
& $0.006$ & \begin{tabular}[t]{@{}c@{}}$0.000$\\[-1mm]{\scriptsize $[-0.000,0.000]$}\end{tabular}
& $0.052$ & \begin{tabular}[t]{@{}c@{}}$0.000$\\[-1mm]{\scriptsize $[-0.000,0.000]$}\end{tabular} \\

\bottomrule
\end{tabular}
\end{adjustbox}
\end{table}

Our model selects substantially fewer predictors for self-efficacy than for latent academic achievement. Two predictors are selected for both outcomes: Strengths and difficulties questionnaire (SDQ) conduct problems \citep{goodman1997strengths} and psychological sense of school membership (sense of belonging at school). Conduct problems are negatively associated with both self-efficacy and latent academic achievement, whereas sense of belonging at school is positively associated with both outcomes. These findings suggest that behavioural difficulties and a sense of belonging at school are important predictors of both self-efficacy and academic achievement.

For self-efficacy, the only additional selected predictors are having at least one good friend and participation in extracurricular activities, both of which have positive estimated effects. In contrast, latent academic achievement is associated with a broader set of predictors spanning school attendance, health, neurodevelopmental conditions, parental education and teacher communication. Taken together, these findings suggest that the direct predictors of self-efficacy, after accounting for latent academic achievement, are relatively concentrated among proximal social and school-connectedness factors. By contrast, latent academic achievement appears to be embedded in a broader ecology of educational participation, health, neurodevelopmental and family-background factors. This pattern is consistent with social-cognitive accounts of self-efficacy as a relatively proximal, domain-specific belief about capability, shaped by interpreted mastery experience and self-regulatory processes \citep{WoodBandura1989Ability,bandura1997self}, while academic achievement is more commonly treated as a cumulative outcome shaped by multiple individual, family, school and health-related influences \citep{hattie2009visible}.

\subsection{The relationship between parental academic assessment and self-efficacy}

A key finding is that latent academic achievement, underlying the observed parental assessment of academic performance, remains positively associated with self-efficacy after adjusting for a broad range of child, family, school and health-related covariates. The posterior mean of the dependence parameter is $\hat{\delta}_{12}=0.070$, with a 95\% credible interval of $[0.048,0.092]$.

\begin{figure}[h]
    \centering
    \begin{subfigure}{0.48\textwidth}
        \centering
        \includegraphics[width=\textwidth]{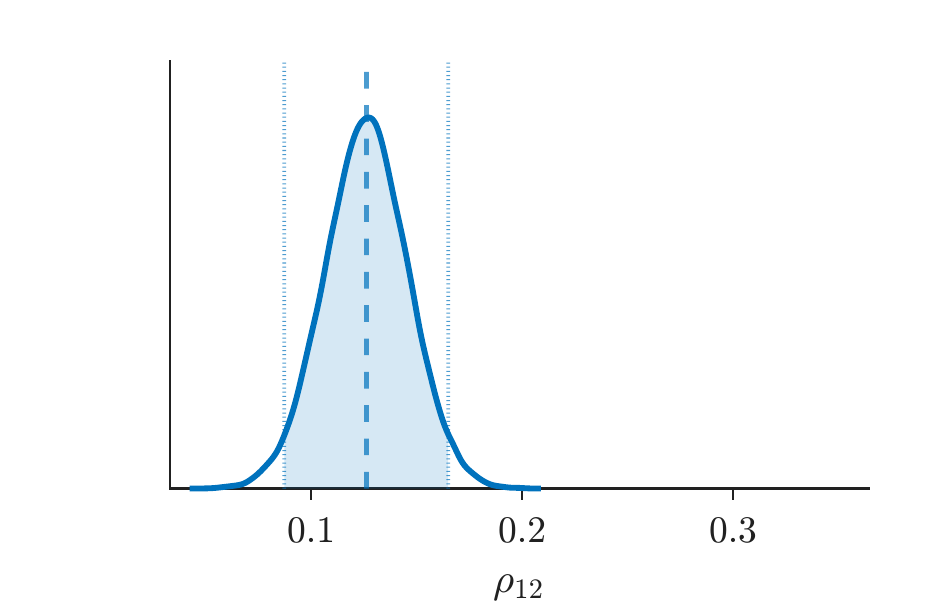}
        \caption{Adjusted model (with covariates).}
    \end{subfigure}
    \hfill
    \begin{subfigure}{0.48\textwidth}
        \centering
        \includegraphics[width=\textwidth]{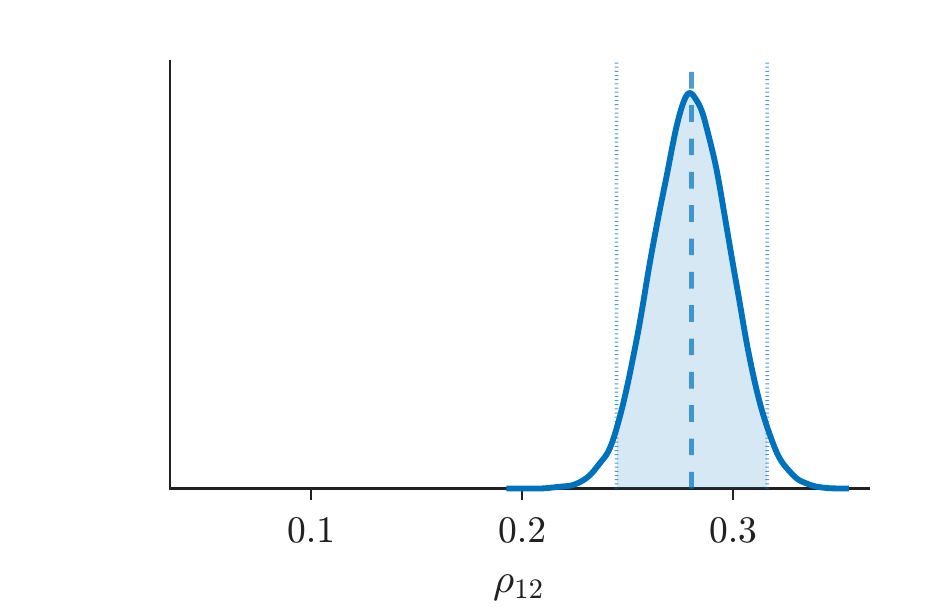}
        \caption{Unadjusted model (without covariates).}
    \end{subfigure}
    \caption{Posterior distributions of the conditional correlation $\rho_{12}$ in \eqref{eq:conditional_corr} for the adjusted model (a) and the unadjusted model (b). The shaded areas correspond to the 95\% credible intervals, and the vertical lines denote the posterior mean estimates.}
    \label{fig:cond_correlation}
\end{figure}

To investigate the extent to which this association is explained by observed covariates, we also fit a version of the model that excludes all covariates. In the unadjusted model, the posterior mean of the dependence parameter increases to $\hat{\delta}_{12}=0.190$, with a 95\% credible interval of $[0.165,0.216]$. To facilitate interpretation, the estimated dependence parameters are converted to the conditional correlation using \eqref{eq:conditional_corr}. Figure \ref{fig:cond_correlation} shows the posterior distributions of the conditional correlation between latent academic achievement and self-efficacy for both the adjusted and unadjusted models. In the unadjusted model, the posterior mean conditional correlation is $0.280$, with a 95\% credible interval of $[0.245,0.316]$. After adjusting for the observed covariates, the corresponding posterior mean conditional correlation decreases to $0.126$, with a 95\% credible interval of $[0.087,0.165]$, representing a reduction of approximately 55\%. This indicates that a substantial proportion of the association between latent academic achievement and self-efficacy is explained by observed child, family, school and health-related characteristics. Nevertheless, a clear positive association remains after adjustment. This finding is consistent with, but does not by itself prove, the interpretation that perceived academic mastery is relatively proximal to self-efficacy, whereas academic achievement is a more cumulative outcome shaped by multiple individual, family, school and health-related influences.

Since latent academic achievement is inferred from parents' assessments rather than objective test scores, the remaining association suggests that parental assessments capture an important subjective sense of academic mastery that remains positively associated with children's self-efficacy, even after accounting for a rich set of observed covariates. Policy-relevant implications of this finding are discussed in the next subsection.

\subsection{Policy-relevant implications}

The results from the previous subsection have policy relevance because latent academic achievement, inferred from parents' assessments, remains positively associated with self-efficacy after adjustment for child, family, school and health-related covariates. This suggests that parents' assessments may capture an interpreted sense of academic mastery relevant to children's confidence in their own capabilities, rather than merely duplicating objective measures of performance. Latent academic achievement measured in this way may therefore represent an important pathway through which educational experiences shape self-efficacy.

\begin{figure}[h]
    \centering
    \begin{subfigure}{0.48\textwidth}
        \centering
        \includegraphics[width=\textwidth]{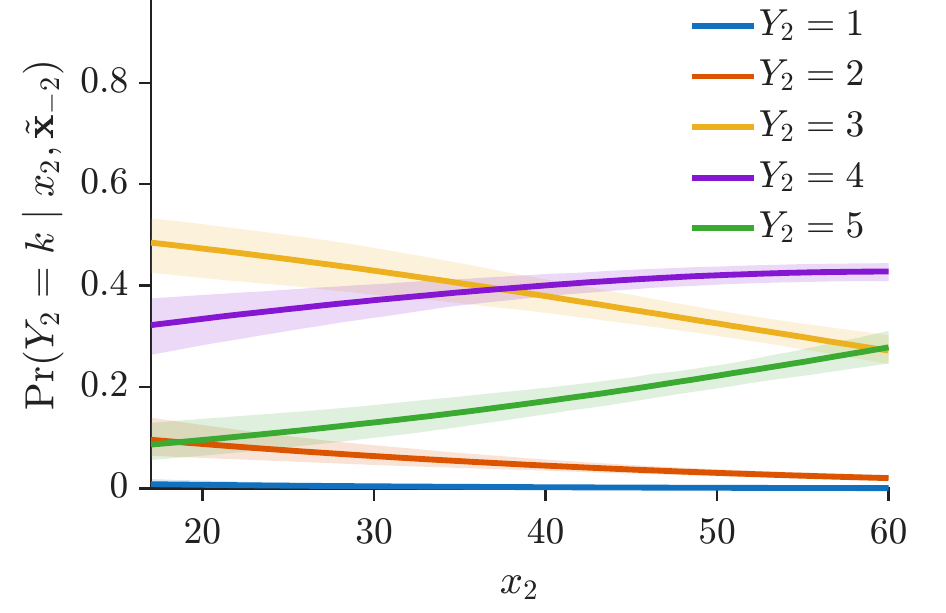}
        \caption{Sense of belonging at school.}
        \label{fig:pp_school_membership}
    \end{subfigure}
    \hfill
    \begin{subfigure}{0.48\textwidth}
        \centering
        \includegraphics[width=\textwidth]{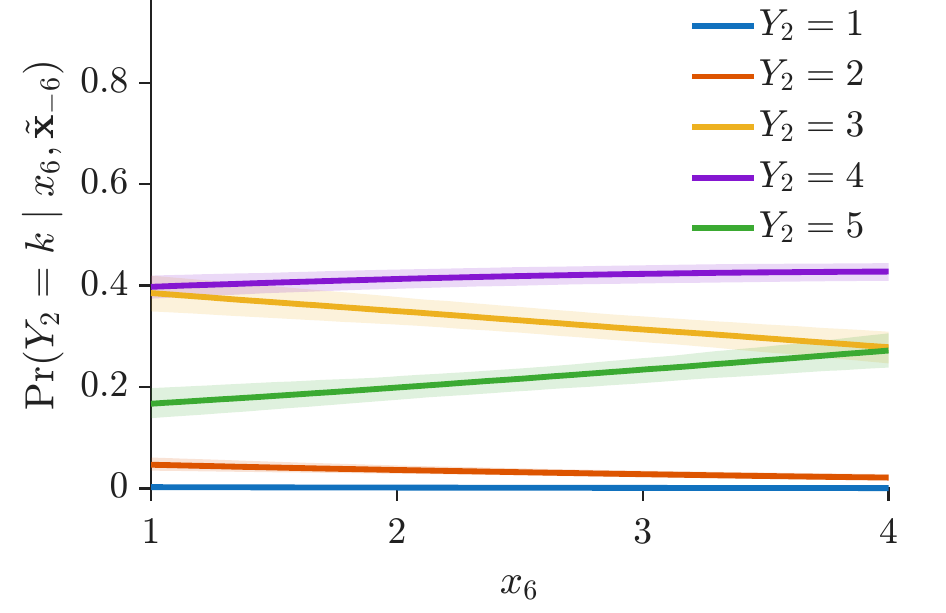}
        \caption{Teacher communication.}
        \label{fig:pp_teacher_communication}
    \end{subfigure}
    \caption{Posterior predictive probabilities for the parental assessment of academic achievement as a function of selected policy-relevant covariates. All remaining covariates, denoted by $\tilde{\mathbf{x}}_{-j}$, are fixed at their sample median values, where $j\in\{2,6\}$ corresponds to the covariate shown in the panel. Panel (a) varies sense of belonging at school (psychological sense of school membership) ($x_2$), while panel (b) varies teacher communication ($x_6$). The curves show posterior median probabilities for each academic achievement category, together with pointwise 95\% credible intervals.}
    \label{fig:pp_policy_covariates}
\end{figure}

To highlight the potential policy implications of these findings, we examine posterior predictive probabilities for selected policy-relevant covariates identified as important by the Bayesian variable selection procedure. For a selected covariate $x_j$, we evaluate the fitted probabilities of each academic achievement category as $x_j$ varies over its observed range, while the remaining covariates are fixed at their sample median values $\tilde{\mathbf{x}}_{-j}$. Specifically,
\[
\Pr(Y_2=k | x_j,\tilde{\mathbf{x}}_{-j})
=
\Phi\!\left(\xi_k-\mathbf{x}^\top\boldsymbol{\beta}_2\right)
-
\Phi\!\left(\xi_{k-1}-\mathbf{x}^\top\boldsymbol{\beta}_2\right),
\qquad k=1,\ldots,K,
\]
where $\mathbf{x}$ is formed by replacing the $j$th covariate with $x_j$ and fixing all remaining covariates at their sample median values. Figure~\ref{fig:pp_policy_covariates} shows the posterior predictive probabilities for sense of belonging at school (left panel) and teacher communication (right panel). Both variables were identified by the Bayesian variable selection procedure as important predictors of academic achievement and represent aspects of the educational environment that may be influenced through policy and educational practice. As sense of belonging at school and teacher communication improve, the posterior predictive probabilities shift towards higher categories of academic achievement. In particular, the probabilities of the two highest categories increase, whereas the probability of the middle category decreases and those of the two lowest categories remain largely unchanged.

The implication of these findings is not that parental assessments should replace objective measures such as NAPLAN\footnote{National Assessment Program--Literacy and Numeracy (NAPLAN) is an annual set of standardised tests for Australian students in Years 3, 5, 7, and 9.}, nor that parents should be encouraged to overstate children’s performance. Rather, the results suggest that policy should attend to the educational conditions through which children and parents recognise genuine academic progress. This includes constructive academic feedback, effective parent–school communication and supportive school environments that help students interpret setbacks as opportunities for improvement while maintaining a strong sense of academic capability. This is also consistent with \citet{honicke2023self}, who argue that targeting self-efficacy directly may not produce the largest educational gains; our results suggest that strengthening the experiences through which academic mastery is developed and recognised may be a more effective indirect pathway.

\section{Conclusion and future research}\label{sec:conclusion_future_reseach}

We propose a Bayesian variable selection methodology for jointly modelling a continuous and an ordinal response through a latent Gaussian representation, motivated by evidence from the educational literature that academic achievement influences self-efficacy. By modelling self-efficacy conditionally on latent academic achievement, the proposed formulation provides a direct and interpretable characterisation of how latent academic achievement relates to self-efficacy after adjusting for observed covariates. At the same time, this conditional parameterisation facilitates analytical marginalisation of the regression coefficients when updating the variable inclusion indicators, leading to a tailored partially collapsed Gibbs sampler for posterior inference. The resulting sampler yields substantial computational benefits relative to the more general Gaussian copula regression approach in \citet{alexopoulos2021bayesian}. Indeed, the simulation study in the supplementary material demonstrates improved variable-selection performance, substantially larger effective sample sizes for a fixed number of Markov chain Monte Carlo iterations, and reduced computational time. However, the Gaussian copula regression framework remains considerably more general and applicable to a broad class of mixed-response problems. The proposed methodology nevertheless illustrates the substantial gains that can be achieved by tailoring the inferential strategy to exploit the application-specific conditional structure suggested by the motivating scientific question.

Applied to the LSAC dataset, the proposed methodology provides answers to several important applied questions. Consistent with the directional evidence motivating the model, latent academic achievement remains positively associated with self-efficacy after adjusting for a broad range of child, family, school and health-related characteristics, although this association is substantially attenuated after adjustment. The variable-selection results reveal a marked asymmetry in the determinants of the two outcomes: self-efficacy is primarily associated with children's social connectedness and sense of belonging at school, whereas latent academic achievement is linked to a broader set of educational, health and family-related factors. This pattern is consistent with theoretical accounts of self-efficacy as a proximal construct shaped by interpreted mastery experience \citep{WoodBandura1989Ability,bandura1997self}, while academic achievement reflects a wider range of individual, family, school and health influences \citep{hattie2009visible}. The proposed variable-selection framework is well placed to characterise this distinction empirically, and the present analysis is consistent with this perspective. Finally, the policy-related posterior predictive analysis highlights sense of belonging at school and teacher communication as potential pathways through which educational environments may contribute to improved academic achievement and, indirectly, to children's self-efficacy.

Several directions for future research merit attention. First, the proposed framework could be extended to multivariate mixed-response settings involving several continuous, ordinal and other non-Gaussian outcomes. One possible approach is to construct an ordered sequence of conditional models among the responses, thereby retaining the interpretable conditional formulations developed in this paper while accommodating richer dependence structures. Such an extension has the potential to bridge the gap towards the generality of Gaussian copula regression while preserving the computational advantages of analytical marginalisation and efficient partially collapsed Gibbs sampling wherever conditional structure can be exploited. Second, as noted in Section~\ref{subsec:validation_posterior_predictive}, the Gaussian assumption does not capture the pronounced bimodal structure of the observed self-efficacy distribution; replacing it with a finite mixture of Gaussian distributions could provide a more flexible characterisation and warrants further investigation. Finally, the present analysis uses a single wave of the LSAC data. Exploiting the longitudinal structure of the study to model trajectories of academic achievement and self-efficacy over time represents a natural and substantively important extension.

\section*{Acknowledgements}

The authors gratefully acknowledge the support of the Paul Ramsay Foundation through the Thrive: Finishing School Well initiative. This work was also supported by the Commonwealth Scientific and Industrial Research Organisation's (CSIRO) Next Generation Graduates Program, an initiative that provides funding and training support to develop the next generation of technology specialists. Sarah Lee benefited from discussions with the Centre for Education Statistics and Evaluation (CESE) team at the New South Wales Department of Education. The views expressed in this paper are those of the authors and do not necessarily reflect those of the New South Wales Department of Education.

\clearpage

\bibliographystyle{apalike}
\bibliography{ref_jrssa}

\appendix

\renewcommand{\thesection}{\Alph{section}}
\numberwithin{equation}{section}
\renewcommand{\theequation}{\thesection.\arabic{equation}}

\clearpage
\appendix
\renewcommand{\thesection}{S\arabic{section}}

\renewcommand{\thesubsection}{\thesection.\arabic{subsection}}

\renewcommand{\theequation}{S\arabic{section}.\arabic{equation}}

\setcounter{section}{0}
\setcounter{equation}{0}
\setcounter{lemma}{0}
\setcounter{remark}{0}
\setcounter{proposition}{0}
\setcounter{figure}{0}
\setcounter{table}{0}

\renewcommand{\thefigure}{S\arabic{figure}}
\renewcommand{\thetable}{S\arabic{table}}
\renewcommand{\thelemma}{S\arabic{lemma}}
\renewcommand{\theremark}{S\arabic{remark}}
\renewcommand{\theproposition}{S\arabic{proposition}}

\setcounter{page}{1}
\renewcommand{\thepage}{S\arabic{page}}

\section{Derivation of full conditional distributions in the partially collapsed Gibbs sampler}\label{app:derivation_Gibbs}

\subsection{Posterior distribution}

We now derive the posterior distribution corresponding to the model without variable selection in more detail than in Section \ref{subsec:posterior_distribution}.

Let
\begin{align*}
\mathbf{y}_1 &= (y_{11},\dots,y_{1n})^\top, \\
\mathbf{y}_2 &= (y_{21},\dots,y_{2n})^\top,
\end{align*}
and let
\begin{align*}
\mathbf{y}_2^* = (y_{21}^*,\dots,y_{2n}^*)^\top
\end{align*}
denote the corresponding latent variables. The posterior distribution is
\begin{align}
\label{eq:joint_posterior_no_VS}
&p(\mathbf{y}_2^*, \boldsymbol{\beta}_1, \boldsymbol{\beta}_2, \xi_1,\dots,\xi_{K-1}, \delta_{12}, \nu_{11} | \mathbf{y}_1, \mathbf{y}_2)
\nonumber\\
&\propto
p(\mathbf{y}_1,\mathbf{y}_2 | \mathbf{y}_2^*, \boldsymbol{\beta}_1, \boldsymbol{\beta}_2, \xi_1,\dots,\xi_{K-1}, \delta_{12}, \nu_{11})
\nonumber\\
&\quad\times
p(\mathbf{y}_2^*, \boldsymbol{\beta}_1, \boldsymbol{\beta}_2, \xi_1,\dots,\xi_{K-1}, \delta_{12}, \nu_{11}).
\end{align}

The factorisation of the likelihood follows directly from the conditional independence structure encoded in Figure \ref{fig:graphical_model}. Specifically,
\begin{align*}
&p(\mathbf{y}_1,\mathbf{y}_2 | \mathbf{y}_2^*, \boldsymbol{\beta}_1, \boldsymbol{\beta}_2, \xi_1,\dots,\xi_{K-1}, \delta_{12}, \nu_{11})
\\
&=
p(\mathbf{y}_2 | \mathbf{y}_2^*, \xi_1,\dots,\xi_{K-1})
\,
p(\mathbf{y}_1 | \mathbf{y}_2^*, \boldsymbol{\beta}_1, \boldsymbol{\beta}_2, \delta_{12}, \nu_{11})
\\
&=
\prod_{i=1}^n
\Pr(y_{2i} | y_{2i}^*, \xi_1,\dots,\xi_{K-1})
\,
p(y_{1i} | y_{2i}^*, \boldsymbol{\beta}_1, \boldsymbol{\beta}_2, \delta_{12}, \nu_{11})
\\
&=
\prod_{i=1}^n
\mathbf{1}\{\xi_{y_{2i}-1} \le y_{2i}^* \le \xi_{y_{2i}}\}
\,
p(y_{1i} | y_{2i}^*, \boldsymbol{\beta}_1, \boldsymbol{\beta}_2, \delta_{12}, \nu_{11}),
\end{align*}
where the final line follows from the threshold representation of the ordinal response. Moreover, the prior factorises as
\begin{align*}
&p(\mathbf{y}_2^*, \boldsymbol{\beta}_1, \boldsymbol{\beta}_2, \xi_1,\dots,\xi_{K-1}, \delta_{12}, \nu_{11})
\\
&=
p(\boldsymbol{\beta}_1,\delta_{12} | \nu_{11})
\,
p(\nu_{11})
\,
p(\boldsymbol{\beta}_2)
\left(
\prod_{i=1}^n
p(y_{2i}^* | \boldsymbol{\beta}_2)
\right)
\,
p(\xi_1,\dots,\xi_{K-1}),
\end{align*}
where the factorisation follows from the prior specification adopted in Section \ref{subsec:model}.

Substituting these expressions into \eqref{eq:joint_posterior_no_VS} yields
\begin{align}
\label{eq:posterior_all_no_VS}
&p(\mathbf{y}_2^*, \boldsymbol{\beta}_1, \boldsymbol{\beta}_2, \xi_1,\dots,\xi_{K-1}, \delta_{12}, \nu_{11} | \mathbf{y}_1, \mathbf{y}_2)
\nonumber\\
&\propto
\left(
\prod_{i=1}^n
\mathbf{1}\{\xi_{y_{2i}-1} \le y_{2i}^* \le \xi_{y_{2i}}\}
\,
p(y_{1i} | y_{2i}^*, \boldsymbol{\beta}_1, \boldsymbol{\beta}_2, \delta_{12}, \nu_{11})
\right)
\nonumber\\
&\quad\times
p(\boldsymbol{\beta}_1,\delta_{12} | \nu_{11})
\,
p(\nu_{11})
\,
p(\boldsymbol{\beta}_2)
\left(
\prod_{i=1}^n
p(y_{2i}^* | \boldsymbol{\beta}_2)
\right)
\,
p(\xi_1,\dots,\xi_{K-1}).
\end{align}
We next derive the full conditional and marginal full conditional distributions used in the partially collapsed Gibbs sampler. In what follows, ``rest'' denotes the collection of variables conditioned upon in each update, as specified by Algorithm \ref{alg:partially_collapsed_Gibbs}.

\subsection{Step 1: Sampling $\boldsymbol{\gamma}_1| \text{rest}$}

Conditionally on the current values of $\mathbf{y}_2^*$, $\boldsymbol{\beta}_{2,\mathcal{A}_2}$, and $\boldsymbol{\gamma}_2$, define
\begin{align*}
\mathbf{e}_2 = \mathbf{y}_2^* - \mathbf{X}_{\mathcal{A}_2}\boldsymbol{\beta}_{2,\mathcal{A}_2}.
\end{align*}
Then \eqref{eq:conditional_model} can be written as
\begin{align}\label{eq:y_1_with_error}
\mathbf{y}_1
=
\mathbf{X}_{\mathcal{A}_1}\boldsymbol{\beta}_{1,\mathcal{A}_1}
+
\delta_{12}\mathbf{e}_2
+
\mathbf{e}_1,
\qquad
\mathbf{e}_1 \sim \mathcal{N}(\mathbf{0}, \nu_{11}\mathbf{I}_n).
\end{align}
Define the augmented design matrix
\begin{align*}
\widetilde{\mathbf{X}}_{\mathcal{A}_1}
=
\left[
\mathbf{X}_{\mathcal{A}_1}, \mathbf{e}_2
\right],
\end{align*}
and the augmented coefficient vector
\begin{align*}
\widetilde{\boldsymbol{\beta}}_{\mathcal{A}_1}
=
\begin{pmatrix}
\boldsymbol{\beta}_{1,\mathcal{A}_1} \\
\delta_{12}
\end{pmatrix}.
\end{align*}
Then, it follows from \eqref{eq:y_1_with_error} that
\begin{align*}
\mathbf{y}_1 | \widetilde{\boldsymbol{\beta}}_{\mathcal{A}_1}, \nu_{11}, \boldsymbol{\gamma}_1, \text{rest}
\sim
\mathcal{N}\left(
\widetilde{\mathbf{X}}_{\mathcal{A}_1}\widetilde{\boldsymbol{\beta}}_{\mathcal{A}_1},
\nu_{11}\mathbf{I}_n
\right).
\end{align*}
For conjugacy, assume a Zellner--$g$ prior for the non-zero coefficients,
\begin{align}\label{eq:Zellner_beta_1_cond_gamma1}
\widetilde{\boldsymbol{\beta}}_{\mathcal{A}_1} | \nu_{11}
\sim
\mathcal{N}\left(
\mathbf{0},
g_1 \nu_{11}
\left(
\widetilde{\mathbf{X}}_{\mathcal{A}_1}^\top
\widetilde{\mathbf{X}}_{\mathcal{A}_1}
\right)^{-1}
\right),
\end{align}
and the inverse-gamma prior
\begin{align*}
\nu_{11} \sim \mathrm{IG}(a_\nu,b_\nu).
\end{align*}

We seek the collapsed conditional distribution of $\boldsymbol{\gamma}_1$, integrating out
\begin{align*}
(\boldsymbol{\beta}_{1,\mathcal{A}_1}, \delta_{12}, \nu_{11}).
\end{align*}
Thus
\begin{align*}
p(\boldsymbol{\gamma}_1 | \text{rest})
\propto
p(\mathbf{y}_1 | \boldsymbol{\gamma}_1, \text{rest except } \boldsymbol{\beta}_1,\delta_{12},\nu_{11})
\, p(\boldsymbol{\gamma}_1|\pi_1),
\end{align*}
where
\begin{align*}
&p(\mathbf{y}_1 | \boldsymbol{\gamma}_1, \text{rest except } \boldsymbol{\beta}_1,\delta_{12},\nu_{11}) \\
&=
\int
p(\mathbf{y}_1 | \widetilde{\boldsymbol{\beta}}_{\mathcal{A}_1}, \nu_{11}, \boldsymbol{\gamma}_1)
\,
p(\widetilde{\boldsymbol{\beta}}_{\mathcal{A}_1} | \nu_{11}, \boldsymbol{\gamma}_1)
\,
p(\nu_{11})
\,
d\widetilde{\boldsymbol{\beta}}_{\mathcal{A}_1}
\, d\nu_{11}.
\end{align*}
We first integrate out $\widetilde{\boldsymbol{\beta}}_{\mathcal{A}_1}$. Under the Zellner--$g$ prior in \eqref{eq:Zellner_beta_1_cond_gamma1},
\begin{align*}
\mathbf{y}_1 | \nu_{11}, \boldsymbol{\gamma}_1
\sim
\mathcal{N}\left(
\mathbf{0},
\nu_{11}(\mathbf{I}_n + g_1 \mathbf{H}_{\mathcal{A}_1})
\right),
\end{align*}
where
\begin{align*}
\mathbf{H}_{\mathcal{A}_1}
=
\widetilde{\mathbf{X}}_{\mathcal{A}_1}
\left(
\widetilde{\mathbf{X}}_{\mathcal{A}_1}^\top
\widetilde{\mathbf{X}}_{\mathcal{A}_1}
\right)^{-1}
\widetilde{\mathbf{X}}_{\mathcal{A}_1}^\top
\end{align*}
is the hat matrix. Hence,
\begin{align*}
p(\mathbf{y}_1 | \nu_{11}, \boldsymbol{\gamma}_1)
&\propto
\nu_{11}^{-n/2}
|\mathbf{I}_n + g_1 \mathbf{H}_{\mathcal{A}_1}|^{-1/2} \\
&\quad \times
\exp\left(
-\frac{1}{2\nu_{11}}
\mathbf{y}_1^\top
(\mathbf{I}_n + g_1 \mathbf{H}_{\mathcal{A}_1})^{-1}
\mathbf{y}_1
\right).
\end{align*}
Using the identities
\begin{align*}
|\mathbf{I}_n + g_1 \mathbf{H}_{\mathcal{A}_1}|
&=
(1+g_1)^{r_{\boldsymbol{\gamma}_1}}, \\
(\mathbf{I}_n + g_1 \mathbf{H}_{\mathcal{A}_1})^{-1}
&=
\mathbf{I}_n - \frac{g_1}{1+g_1}\mathbf{H}_{\mathcal{A}_1},
\end{align*}
where
\begin{align*}
r_{\boldsymbol{\gamma}_1}
=
\mathrm{rank}(\widetilde{\mathbf{X}}_{\mathcal{A}_1}),
\end{align*}
we obtain
\begin{align*}
p(\mathbf{y}_1 | \nu_{11}, \boldsymbol{\gamma}_1)
&\propto
\nu_{11}^{-n/2}
(1+g_1)^{-r_{\boldsymbol{\gamma}_1}/2} \\
&\quad \times
\exp\left(
-\frac{1}{2\nu_{11}}
\mathbf{y}_1^\top
\mathbf{Q}_{\mathcal{A}_1}
\mathbf{y}_1
\right),
\end{align*}
with
\begin{align*}
\mathbf{Q}_{\mathcal{A}_1}
=
\mathbf{I}_n - \frac{g_1}{1+g_1}\mathbf{H}_{\mathcal{A}_1}.
\end{align*}

Next, integrating out $\nu_{11} \sim \mathrm{IG}(a_\nu,b_\nu)$ gives
\begin{align*}
&p(\mathbf{y}_1 | \boldsymbol{\gamma}_1, \text{rest except } \boldsymbol{\beta}_1,\delta_{12},\nu_{11}) \\
&\propto
(1+g_1)^{-r_{\boldsymbol{\gamma}_1}/2}
\frac{\Gamma(a_\nu + n/2)}{\Gamma(a_\nu)}
\frac{b_\nu^{a_\nu}}
{\left(
b_\nu + \frac{1}{2}\mathbf{y}_1^\top \mathbf{Q}_{\mathcal{A}_1}\mathbf{y}_1
\right)^{a_\nu+n/2}}.
\end{align*}
Therefore, when updating one coordinate $\gamma_{1j}$ at a time in a Gibbs-within-partially-collapsed Gibbs scheme, we compare the two values
\begin{align*}
\gamma_{1j}=0
\qquad \text{and} \qquad
\gamma_{1j}=1,
\end{align*}
while holding the remaining components fixed, and compute
\begin{align*}
\Pr(\gamma_{1j}=1 | \text{rest})
=
\frac{
p(\mathbf{y}_1 | \gamma_{1j}=1,\boldsymbol{\gamma}_{1,-j},\text{rest}) \, \pi_1
}{
p(\mathbf{y}_1 | \gamma_{1j}=1,\boldsymbol{\gamma}_{1,-j},\text{rest}) \, \pi_1
+
p(\mathbf{y}_1 | \gamma_{1j}=0,\boldsymbol{\gamma}_{1,-j},\text{rest}) \, (1-\pi_1)
}.
\end{align*}

\subsection{Step 2: Sampling $(\boldsymbol{\beta}_{1,\mathcal{A}_1}, \delta_{12}) | \text{rest}$}
Recall the conditional model in \eqref{eq:y_1_with_error}, 
\begin{align}
\label{eq:y1_reg}
\mathbf{y}_1
=
\mathbf{X}_{\mathcal{A}_1}\boldsymbol{\beta}_{1,\mathcal{A}_1}
+
\delta_{12}\mathbf{e}_2
+
\mathbf{e}_1,
\end{align}
where
\begin{align*}
\mathbf{e}_1 \sim \mathcal{N}(\mathbf{0}, \nu_{11}\mathbf{I}_n),
\end{align*}
and
\begin{align*}
\mathbf{e}_2
=
\mathbf{y}_2^* - \mathbf{X}_{\mathcal{A}_2}\boldsymbol{\beta}_{2,\mathcal{A}_2}.
\end{align*}
Thus, conditional on the current values of $\mathbf{y}_2^*$, $\boldsymbol{\beta}_{2,\mathcal{A}_2}$, $\boldsymbol{\gamma}_1$, $\boldsymbol{\gamma}_2$, and $\nu_{11}$, \eqref{eq:y1_reg} is a Gaussian linear regression in $\boldsymbol{\beta}_{1,\mathcal{A}_1}$ and $\delta_{12}$. Define the augmented design matrix
\begin{align*}
\widetilde{\mathbf{X}}
=
\left[
\mathbf{X}_{\mathcal{A}_1},
\mathbf{e}_2
\right],
\end{align*}
and the augmented coefficient vector
\begin{align*}
\widetilde{\boldsymbol{\beta}}
=
\begin{pmatrix}
\boldsymbol{\beta}_{1,\mathcal{A}_1} \\
\delta_{12}
\end{pmatrix}.
\end{align*}
Then, from \eqref{eq:y1_reg} it follows that
\begin{align*}
\mathbf{y}_1 | \widetilde{\boldsymbol{\beta}}, \nu_{11}, \boldsymbol{\gamma}_1, \boldsymbol{\gamma}_2, \mathbf{y}_2^*, \boldsymbol{\beta}_{2,\mathcal{A}_2}
\sim
\mathcal{N}\!\left(
\widetilde{\mathbf{X}}\widetilde{\boldsymbol{\beta}},
\nu_{11}\mathbf{I}_n
\right).
\end{align*}

For conjugacy, we assume a Zellner--$g$ prior for the non-zero coefficients of $\widetilde{\boldsymbol{\beta}}$, i.e.\ for the active elements of $\boldsymbol{\beta}_1$ together with $\delta_{12}$
\begin{align*}
\widetilde{\boldsymbol{\beta}} | \nu_{11}
\sim
\mathcal{N}\!\left(
\mathbf{0},
g_1 \nu_{11}
(\widetilde{\mathbf{X}}^\top \widetilde{\mathbf{X}})^{-1}
\right).
\end{align*}
Let
\begin{align*}
\widehat{\widetilde{\boldsymbol{\beta}}}_{\mathrm{OLS}}
=
(\widetilde{\mathbf{X}}^\top \widetilde{\mathbf{X}})^{-1}
\widetilde{\mathbf{X}}^\top \mathbf{y}_1
\end{align*}
denote the ordinary least squares estimator based on the active covariates in the augmented design matrix $\widetilde{\mathbf{X}} = [\mathbf{X}_{\mathcal{A}_1}, \mathbf{e}_2]$. By conjugacy, it follows that the full conditional posterior is
\begin{align*}
\widetilde{\boldsymbol{\beta}} | \text{rest}
\sim
\mathcal{N}\!\left(
\frac{g_1}{1+g_1}\widehat{\widetilde{\boldsymbol{\beta}}}_{\mathrm{OLS}},
\frac{g_1}{1+g_1}\nu_{11}
(\widetilde{\mathbf{X}}^\top \widetilde{\mathbf{X}})^{-1}
\right),
\end{align*}
and thus
\begin{align*}
(\boldsymbol{\beta}_{1,\mathcal{A}_1}, \delta_{12}) | \text{rest}
\sim
\mathcal{N}\!\left(
\frac{g_1}{1+g_1}\widehat{\widetilde{\boldsymbol{\beta}}}_{\mathrm{OLS}},
\frac{g_1}{1+g_1}\nu_{11}
(\widetilde{\mathbf{X}}^\top \widetilde{\mathbf{X}})^{-1}
\right).
\end{align*}

\subsection{Step 3: Sampling $\pi_1 | \text{rest}$}\label{subsec:pi_1_full_conditional}
The full conditional distribution of $\pi_1$ is given by
\begin{align*}
p(\pi_1 | \text{rest})
&\propto
p(\boldsymbol{\gamma}_1 | \pi_1)\, p(\pi_1).
\end{align*}
Using the Bernoulli prior on the inclusion indicators and the beta prior on $\pi_1$, we obtain
\begin{align*}
p(\pi_1 | \text{rest})
&\propto
\prod_{j=1}^p \pi_1^{\gamma_{1j}} (1-\pi_1)^{1-\gamma_{1j}}
\cdot
\pi_1^{a_{\pi,1}-1} (1-\pi_1)^{b_{\pi,1}-1} \\
&=
\pi_1^{a_{\pi,1} + \sum_{j=1}^p \gamma_{1j} - 1}
(1-\pi_1)^{b_{\pi,1} + p - \sum_{j=1}^p \gamma_{1j} - 1}.
\end{align*}
Hence, by conjugacy, it follows that
\begin{align*}
\pi_1 | \text{rest}
\sim
\mathrm{Beta}\left(
a_{\pi,1} + \sum_{j=1}^p \gamma_{1j},
\;
b_{\pi,1} + p - \sum_{j=1}^p \gamma_{1j}
\right).
\end{align*}

\subsection{Step 4: Sampling $\boldsymbol{\gamma}_2|\text{rest}$}
To update $\boldsymbol{\gamma}_2$, we integrate out $\boldsymbol{\beta}_{2,\mathcal{A}_2}$. The collapsed full conditional is
\begin{align}\label{eq:gamma_2_marginalised_conditional}
p(\boldsymbol{\gamma}_2 | \mathbf{y}_1, \mathbf{y}_2^*, \text{rest})
&\propto
p(\mathbf{y}_1, \mathbf{y}_2^* | \boldsymbol{\gamma}_2, \text{rest}) \, p(\boldsymbol{\gamma}_2|\pi_2) \nonumber \\
&=
p(\mathbf{y}_1 | \mathbf{y}_2^*, \boldsymbol{\gamma}_2, \text{rest})
\, p(\mathbf{y}_2^* | \boldsymbol{\gamma}_2)
\, p(\boldsymbol{\gamma}_2|\pi_2).
\end{align}

The first marginal likelihood factor in \eqref{eq:gamma_2_marginalised_conditional},
\begin{align*}
p(\mathbf{y}_1 | \mathbf{y}_2^*, \boldsymbol{\gamma}_2, \text{rest}),
\end{align*}
is obtained by integrating out $\boldsymbol{\beta}_{2,\mathcal{A}_2}$ from \eqref{eq:y1_reg} conditional on $\mathbf{y}_2^*$. First note that
\begin{align*}
\mathbf{y}_1
&=
\mathbf{X}_{\mathcal{A}_1}\boldsymbol{\beta}_{1,\mathcal{A}_1}
+
\delta_{12}
\left(
\mathbf{y}_2^*
-
\mathbf{X}_{\mathcal{A}_2}
\boldsymbol{\beta}_{2,\mathcal{A}_2}
\right)
+
\mathbf{e}_1,
\qquad
\mathbf{e}_1\sim \mathcal{N}(\mathbf{0},\nu_{11}\mathbf{I}_n).
\end{align*}
Conditional on $\mathbf{y}_2^*$ and $\boldsymbol{\gamma}_2$, the posterior distribution of
$\boldsymbol{\beta}_{2,\mathcal{A}_2}$ under the latent Gaussian regression is
\begin{align*}
\boldsymbol{\beta}_{2,\mathcal{A}_2}
|
\mathbf{y}_2^*
\sim
\mathcal{N}
\left(
\frac{g_2}{1+g_2}
\widehat{\boldsymbol{\beta}}_{2,\mathrm{OLS}},
\;
\frac{g_2}{1+g_2}
(\mathbf{X}_{\mathcal{A}_2}^{\top}
\mathbf{X}_{\mathcal{A}_2})^{-1}
\right),
\end{align*}
where
\begin{align*}
\widehat{\boldsymbol{\beta}}_{2,\mathrm{OLS}}
=
(\mathbf{X}_{\mathcal{A}_2}^{\top}
\mathbf{X}_{\mathcal{A}_2})^{-1}
\mathbf{X}_{\mathcal{A}_2}^{\top}\mathbf{y}_2^*.
\end{align*}
Therefore, after integrating out
$\boldsymbol{\beta}_{2,\mathcal{A}_2}$,
\begin{align*}
\mathbf{y}_1
|
\mathbf{y}_2^*,\boldsymbol{\gamma}_2,\text{rest}
\sim
\mathcal{N}
\left(
\boldsymbol{\mu}_{1|\mathbf{y}_2^*},
\boldsymbol{\Sigma}_{1|\mathbf{y}_2^*}
\right),
\end{align*}
with
\begin{align*}
\boldsymbol{\mu}_{1|\mathbf{y}_2^*}
&=
\mathbf{X}_{\mathcal{A}_1}\boldsymbol{\beta}_{1,\mathcal{A}_1}
+
\delta_{12}
\left(
\mathbf{I}_n
-
\frac{g_2}{1+g_2}\mathbf{H}_2
\right)
\mathbf{y}_2^*,\\
\boldsymbol{\Sigma}_{1|\mathbf{y}_2^*}
&=
\nu_{11}\mathbf{I}_n
+
\delta_{12}^2
\frac{g_2}{1+g_2}\mathbf{H}_2,
\end{align*}
where
\begin{align*}
\mathbf{H}_2
=
\mathbf{X}_{\mathcal{A}_2}
(\mathbf{X}_{\mathcal{A}_2}^{\top}
\mathbf{X}_{\mathcal{A}_2})^{-1}
\mathbf{X}_{\mathcal{A}_2}^{\top}.
\end{align*}
Thus
\begin{align*}
\log p(\mathbf{y}_1|\mathbf{y}_2^*,\boldsymbol{\gamma}_2,\text{rest})
&=
-\frac n2\log(2\pi)
-\frac12\log|\boldsymbol{\Sigma}_{1|\mathbf{y}_2^*}| \\
&\quad
-\frac12
(\mathbf{y}_1-\boldsymbol{\mu}_{1|\mathbf{y}_2^*})^\top
\boldsymbol{\Sigma}_{1|\mathbf{y}_2^*}^{-1}
(\mathbf{y}_1-\boldsymbol{\mu}_{1|\mathbf{y}_2^*}).
\end{align*}

The second ``marginal likelihood'' in \eqref{eq:gamma_2_marginalised_conditional} (with the latent $\mathbf{y}_2^*$ as ``data'') factor comes from integrating out $\boldsymbol{\beta}_{2,\mathcal{A}_2}$ from the latent equation in \eqref{eq:latent_equation}, i.e.\
\begin{align*}
\mathbf{y}_2^*
=
\mathbf{X}_{\mathcal{A}_2}\boldsymbol{\beta}_{2,\mathcal{A}_2}
+
\mathbf{e}_2,
\qquad
\mathbf{e}_2 \sim \mathcal{N}(\mathbf{0}, \mathbf{I}_n),
\end{align*}
with respect to the Zellner--$g$ prior
\begin{align*}
\boldsymbol{\beta}_{2,\mathcal{A}_2} 
\sim
\mathcal{N}\left(
\mathbf{0},
g_2
\left(
\mathbf{X}_{\mathcal{A}_2}^\top \mathbf{X}_{\mathcal{A}_2}
\right)^{-1}
\right).
\end{align*}
This gives 
\begin{align*}
\mathbf{y}_2^* | \boldsymbol{\gamma}_2
\sim
\mathcal{N}\left(
\mathbf{0},
\mathbf{I}_n + g_2 \mathbf{H}_2
\right),
\end{align*}
where
\begin{align*}
\mathbf{H}_2
=
\mathbf{X}_{\mathcal{A}_2}
\left(
\mathbf{X}_{\mathcal{A}_2}^\top \mathbf{X}_{\mathcal{A}_2}
\right)^{-1}
\mathbf{X}_{\mathcal{A}_2}^\top.
\end{align*}
Hence, the log-marginal likelihood is
\begin{align*}
\log p(\mathbf{y}_2^* | \boldsymbol{\gamma}_2)
&=
-\frac n2 \log(2\pi)
-\frac 12 \log |\mathbf{I}_n + g_2 \mathbf{H}_2| \\
&\quad
-\frac 12 (\mathbf{y}_2^*)^\top
(\mathbf{I}_n + g_2 \mathbf{H}_2)^{-1}
\mathbf{y}_2^*.
\end{align*}
Using the identities
\begin{align*}
|\mathbf{I}_n + g_2 \mathbf{H}_2| &= (1+g_2)^r, \\
(\mathbf{I}_n + g_2 \mathbf{H}_2)^{-1}
&=
\mathbf{I}_n - \frac{g_2}{1+g_2}\mathbf{H}_2,
\end{align*}
where
\begin{align*}
r = \mathrm{rank}(\mathbf{X}_{\mathcal{A}_2}),
\end{align*}
we obtain
\begin{align*}
\log p(\mathbf{y}_2^* | \boldsymbol{\gamma}_2)
&=
-\frac n2 \log(2\pi)
-\frac r2 \log(1+g_2) \\
&\quad
-\frac 12 (\mathbf{y}_2^*)^\top
\left(
\mathbf{I}_n - \frac{g_2}{1+g_2}\mathbf{H}_2
\right)
\mathbf{y}_2^*.
\end{align*}

Therefore, when updating one coordinate $\gamma_{2j}$ at a time in a Gibbs-within-partially-collapsed Gibbs scheme, the Bernoulli probability is based on
\begin{align*}
\log p(\mathbf{y}_1 | \mathbf{y}_2^*, \boldsymbol{\gamma}_2, \text{rest})
+
\log p(\mathbf{y}_2^* | \boldsymbol{\gamma}_2)
+
\log p(\boldsymbol{\gamma}_2 | \pi_2),
\end{align*}
where
\begin{align*}
\log p(\boldsymbol{\gamma}_2 | \pi_2)
=
\sum_{l=1}^p
\left[
\gamma_{2l}\log \pi_2
+
(1-\gamma_{2l})\log(1-\pi_2)
\right].
\end{align*}

\subsection{Step 5: Sampling $\boldsymbol{\beta}_{2,\mathcal{A}_2} | \text{rest}$}
We sample the active coefficients $\boldsymbol{\beta}_{2,\mathcal{A}_2}$ conditional on the current values of the remaining parameters. 

Recall the latent equation from \eqref{eq:latent_equation} in vector form, i.e.
\begin{align}\label{eq:intermediate_likelihood}
\mathbf{y}_2^*
=
\mathbf{X}_{\mathcal{A}_2}\boldsymbol{\beta}_{2,\mathcal{A}_2}
+
\mathbf{e}_2,
\qquad
\mathbf{e}_2 \sim \mathcal{N}(\mathbf{0}, \mathbf{I}_n).
\end{align}
Recall also the equation in \eqref{eq:conditional_model} in vector form, given in \eqref{eq:y_1_with_error}, i.e.\
\begin{align*}
\mathbf{y}_1
=
\mathbf{X}_{\mathcal{A}_1}\boldsymbol{\beta}_{1,\mathcal{A}_1}
+
\delta_{12}\mathbf{e}_2
+
\mathbf{e}_1,
\qquad
\mathbf{e}_1 \sim \mathcal{N}(\mathbf{0}, \nu_{11}\mathbf{I}_n).
\end{align*}
Substituting
\begin{align*}
\mathbf{e}_2
=
\mathbf{y}_2^* - \mathbf{X}_{\mathcal{A}_2}\boldsymbol{\beta}_{2,\mathcal{A}_2}
\end{align*}
gives
\begin{align}\label{eq:y_1_with_both_errors}
\mathbf{y}_1
=
\mathbf{X}_{\mathcal{A}_1}\boldsymbol{\beta}_{1,\mathcal{A}_1}
+
\delta_{12}
\left(
\mathbf{y}_2^* - \mathbf{X}_{\mathcal{A}_2}\boldsymbol{\beta}_{2,\mathcal{A}_2}
\right)
+
\mathbf{e}_1.
\end{align}
Define the pseudo-data
\begin{align*}
\mathbf{z}^{(1)}
=
\mathbf{y}_1
-
\mathbf{X}_{\mathcal{A}_1}\boldsymbol{\beta}_{1,\mathcal{A}_1}
-
\delta_{12}\mathbf{y}_2^*.
\end{align*}
Then we can write \eqref{eq:y_1_with_both_errors} as
\begin{align}\label{eq:pseudo_data_model}
\mathbf{z}^{(1)}
=
-\delta_{12}\mathbf{X}_{\mathcal{A}_2}\boldsymbol{\beta}_{2,\mathcal{A}_2}
+
\mathbf{e}_1.
\end{align}

We derive the posterior for the non-zero coefficient  $\boldsymbol{\beta}_{2,\mathcal{A}_2}$ in two steps. First, for conjugacy, we assume a Zellner--$g$ prior for the non-zero coefficients in \eqref{eq:intermediate_likelihood}, i.e.\
\begin{align*}
\boldsymbol{\beta}_{2,\mathcal{A}_2} 
\sim
\mathcal{N}\left(
\mathbf{0},
g_2
\left(
\mathbf{X}_{\mathcal{A}_2}^\top \mathbf{X}_{\mathcal{A}_2}
\right)^{-1}
\right).
\end{align*}
Combining this prior with the (latent data) likelihood in \eqref{eq:intermediate_likelihood}, i.e.
\begin{align*}
\mathbf{y}_2^* | \boldsymbol{\beta}_{2,\mathcal{A}_2}
\sim
\mathcal{N}\left(
\mathbf{X}_{\mathcal{A}_2}\boldsymbol{\beta}_{2,\mathcal{A}_2},
\mathbf{I}_n
\right)
\end{align*}
gives the intermediate posterior
\begin{align*}
\boldsymbol{\beta}_{2,\mathcal{A}_2} | \mathbf{y}_2^*
\sim
\mathcal{N}\left(
g_2' \widehat{\boldsymbol{\beta}}_2^{(2)},
g_2'
\left(
\mathbf{X}_{\mathcal{A}_2}^\top \mathbf{X}_{\mathcal{A}_2}
\right)^{-1}
\right),
\end{align*}
where
\begin{align*}
g_2' = \frac{g_2}{1+g_2},
\qquad
\widehat{\boldsymbol{\beta}}_2^{(2)}
=
\left(
\mathbf{X}_{\mathcal{A}_2}^\top \mathbf{X}_{\mathcal{A}_2}
\right)^{-1}
\mathbf{X}_{\mathcal{A}_2}^\top \mathbf{y}_2^*.
\end{align*}
Next, the intermediate posterior is used as a prior for the model in \eqref{eq:pseudo_data_model}. First, denote the least squares estimator for the non-zero coefficient of $\boldsymbol{\beta}_{2,\mathcal{A}_2}$ of model \eqref{eq:pseudo_data_model} by 
\begin{align*}
\widehat{\boldsymbol{\beta}}_2^{(3)}
=
-\frac{1}{\delta_{12}}
\left(
\mathbf{X}_{\mathcal{A}_2}^\top \mathbf{X}_{\mathcal{A}_2}
\right)^{-1}
\mathbf{X}_{\mathcal{A}_2}^\top \mathbf{z}^{(1)}.
\end{align*}
Combining the Gaussian intermediate posterior with the Gaussian likelihood in \eqref{eq:pseudo_data_model} gives, by conjugacy, \begin{align*}
\boldsymbol{\beta}_{2,\mathcal{A}_2} | \text{rest}
\sim
\mathcal{N}\left(
w_{\mathrm{prior}}\, g_2' \widehat{\boldsymbol{\beta}}_2^{(2)}
+
w_{\mathrm{data}}\, \widehat{\boldsymbol{\beta}}_2^{(3)},
\;
\left(
\frac{1}{g_2'} + \frac{\delta_{12}^2}{\nu_{11}}
\right)^{-1}
\left(
\mathbf{X}_{\mathcal{A}_2}^\top \mathbf{X}_{\mathcal{A}_2}
\right)^{-1}
\right),
\end{align*}
where
\begin{align*}
w_{\mathrm{prior}}
&=
\frac{1/g_2'}{1/g_2' + \delta_{12}^2/\nu_{11}}, \\
w_{\mathrm{data}}
&=
\frac{\delta_{12}^2/\nu_{11}}{1/g_2' + \delta_{12}^2/\nu_{11}}.
\end{align*}

\subsection{Step 6: Sampling $\pi_2 | \text{rest}$}

The full conditional distribution of $\pi_2$ is obtained analogously to that of $\pi_1$ in Section \ref{subsec:pi_1_full_conditional}, yielding
\begin{align*}
\pi_2 | \text{rest}
\sim
\mathrm{Beta}\left(
a_{\pi,2} + \sum_{j=1}^p \gamma_{2j},
\;
b_{\pi,2} + p - \sum_{j=1}^p \gamma_{2j}
\right).
\end{align*}

\subsection{Step 7: Sampling $\mathbf{y}_2^* | \text{rest}$}

From the posterior in \eqref{eq:posterior_all_no_VS}, the full conditional of $\mathbf{y}_2^*$ is proportional to
\begin{align}\label{eq:full_conditional_y2star}
p(\mathbf{y}_2^* | \text{rest})
&\propto
p(\mathbf{y}_1 | \mathbf{y}_2^*, \boldsymbol{\beta}_1, \boldsymbol{\beta}_2, \delta_{12}, \nu_{11})
\,
p(\mathbf{y}_2^* | \boldsymbol{\beta}_2)
\,
\prod_{i=1}^n
\mathbf{1}\{\xi_{y_{2i}-1} \le y_{2i}^* \le \xi_{y_{2i}}\}.
\end{align}
Using the conditional decomposition in vector form, i.e.\
\begin{align*}
\mathbf{y}_1
&=
\mathbf{X}_{\mathcal{A}_1}\boldsymbol{\beta}_{1,\mathcal{A}_1}
+
\delta_{12}
\left(
\mathbf{y}_2^* - \mathbf{X}_{\mathcal{A}_2}\boldsymbol{\beta}_{2,\mathcal{A}_2}
\right)
+
\mathbf{e}_1,
\qquad
\mathbf{e}_1 \sim \mathcal{N}(\mathbf{0}, \nu_{11}\mathbf{I}_n),
\\
\mathbf{y}_2^*
&=
\mathbf{X}_{\mathcal{A}_2}\boldsymbol{\beta}_{2,\mathcal{A}_2}
+
\mathbf{e}_2,
\qquad
\mathbf{e}_2 \sim \mathcal{N}(\mathbf{0}, \mathbf{I}_n),
\end{align*}
we note that the errors are independent (within and between). Thus
the full conditional in \eqref{eq:full_conditional_y2star} factorises over $i$, so it is enough to consider $y_{2i}^*$.

For a fixed $i$,
\begin{align*}
p(y_{2i}^* | \text{rest})
&\propto
p(y_{1i} | y_{2i}^*, \boldsymbol{\beta}_1, \boldsymbol{\beta}_2, \delta_{12}, \nu_{11})
\,
p(y_{2i}^* | \boldsymbol{\beta}_2)
\,
\mathbf{1}\{\xi_{y_{2i}-1} \le y_{2i}^* \le \xi_{y_{2i}}\}.
\end{align*}
Now,
\begin{align*}
y_{1i} | y_{2i}^*, \text{rest}
&\sim
\mathcal{N}\left(
\mathbf{x}_{i,\mathcal{A}_1}^\top \boldsymbol{\beta}_{1,\mathcal{A}_1}
+
\delta_{12}\bigl(y_{2i}^* - \mathbf{x}_{i,\mathcal{A}_2}^\top \boldsymbol{\beta}_{2,\mathcal{A}_2}\bigr),
\,
\nu_{11}
\right),
\\
y_{2i}^* | \text{rest}
&\sim
\mathcal{N}\left(
\mathbf{x}_{i,\mathcal{A}_2}^\top \boldsymbol{\beta}_{2,\mathcal{A}_2},
\, 1
\right).
\end{align*}
Therefore,
\begin{align*}
p(y_{2i}^* | \text{rest})
&\propto
\exp\left\{
-\frac{1}{2\nu_{11}}
\left(
y_{1i}
-
\mathbf{x}_{i,\mathcal{A}_1}^\top \boldsymbol{\beta}_{1,\mathcal{A}_1}
-
\delta_{12} y_{2i}^*
+
\delta_{12}\mathbf{x}_{i,\mathcal{A}_2}^\top \boldsymbol{\beta}_{2,\mathcal{A}_2}
\right)^2
\right\}
\\
&\quad \times
\exp\left\{
-\frac{1}{2}
\left(
y_{2i}^* - \mathbf{x}_{i,\mathcal{A}_2}^\top \boldsymbol{\beta}_{2,\mathcal{A}_2}
\right)^2
\right\}
\,
\mathbf{1}\{\xi_{y_{2i}-1} \le y_{2i}^* \le \xi_{y_{2i}}\}.
\end{align*}
Collecting the quadratic terms in $y_{2i}^*$, we obtain a truncated normal density on the interval $(\xi_{y_{2i}-1},\,\xi_{y_{2i}}]$, i.e.\
\begin{align*}
y_{2i}^* | \text{rest}
\sim
\mathcal{TN}_{(\xi_{y_{2i}-1},\,\xi_{y_{2i}}]}
\left(
\mu_i^*,
\sigma_i^{2*}
\right),
\end{align*}
with
\begin{align*}
\sigma_i^{2*}
&=
\left(
1 + \frac{\delta_{12}^2}{\nu_{11}}
\right)^{-1}
=
\frac{\nu_{11}}{\nu_{11} + \delta_{12}^2},
\\
\mu_i^*
&=
\sigma_i^{2*}
\left[
\frac{\delta_{12}}{\nu_{11}}
\left(
y_{1i}
-
\mathbf{x}_{i,\mathcal{A}_1}^\top \boldsymbol{\beta}_{1,\mathcal{A}_1}
+
\delta_{12}\mathbf{x}_{i,\mathcal{A}_2}^\top \boldsymbol{\beta}_{2,\mathcal{A}_2}
\right)
+
\mathbf{x}_{i,\mathcal{A}_2}^\top \boldsymbol{\beta}_{2,\mathcal{A}_2}
\right].
\end{align*}
Equivalently, defining
\begin{align*}
w_1 &= \frac{\delta_{12}^2}{\nu_{11}+\delta_{12}^2},
&
w_2 &= \frac{\nu_{11}}{\nu_{11}+\delta_{12}^2},
\\
z_i^{(6)}
&=
y_{1i}
-
\mathbf{x}_{i,\mathcal{A}_1}^\top \boldsymbol{\beta}_{1,\mathcal{A}_1}
+
\delta_{12}\mathbf{x}_{i,\mathcal{A}_2}^\top \boldsymbol{\beta}_{2,\mathcal{A}_2},
\end{align*}
we may also write
\begin{align*}
\mu_i^*
=
w_1 \frac{z_i^{(6)}}{\delta_{12}}
+
w_2 \mathbf{x}_{i,\mathcal{A}_2}^\top \boldsymbol{\beta}_{2,\mathcal{A}_2}.
\end{align*}

\paragraph{Step 8: Sampling $\nu_{11} | \text{rest}$}

To maintain consistency with Step 1, we derive the full conditional of $\nu_{11}$ under the same Zellner--$g$ prior on the augmented coefficient vector
\begin{align*}
\widetilde{\boldsymbol{\beta}}_{\mathcal{A}_1}
=
\begin{pmatrix}
\boldsymbol{\beta}_{1,\mathcal{A}_1} \\
\delta_{12}
\end{pmatrix},
\end{align*}
with augmented design matrix
\begin{align*}
\widetilde{\mathbf{X}}_{\mathcal{A}_1}
=
\left[
\mathbf{X}_{\mathcal{A}_1}\, \mathbf{e}_2
\right],
\end{align*}
where
\begin{align*}
\mathbf{e}_2
=
\mathbf{y}_2^* - \mathbf{X}_{\mathcal{A}_2}\boldsymbol{\beta}_{2,\mathcal{A}_2}.
\end{align*}

Recall that
\begin{align*}
\mathbf{y}_1
=
\widetilde{\mathbf{X}}_{\mathcal{A}_1}
\widetilde{\boldsymbol{\beta}}_{\mathcal{A}_1}
+
\mathbf{e}_1,
\qquad
\mathbf{e}_1 \sim \mathcal{N}(\mathbf{0}, \nu_{11}\mathbf{I}_n).
\end{align*}
Hence, the likelihood as a function of $\nu_{11}$ is
\begin{align*}
&p(\mathbf{y}_1 | \widetilde{\boldsymbol{\beta}}_{\mathcal{A}_1}, \nu_{11}, \text{rest}) \\
&\propto
\nu_{11}^{-n/2}
\exp\left(
-\frac{1}{2\nu_{11}}
\left(
\mathbf{y}_1
-
\widetilde{\mathbf{X}}_{\mathcal{A}_1}
\widetilde{\boldsymbol{\beta}}_{\mathcal{A}_1}
\right)^\top
\left(
\mathbf{y}_1
-
\widetilde{\mathbf{X}}_{\mathcal{A}_1}
\widetilde{\boldsymbol{\beta}}_{\mathcal{A}_1}
\right)
\right).
\end{align*}
The prior from Step 1 is
\begin{align*}
\widetilde{\boldsymbol{\beta}}_{\mathcal{A}_1} | \nu_{11}
\sim
\mathcal{N}\left(
\mathbf{0},
g_1 \nu_{11}
\left(
\widetilde{\mathbf{X}}_{\mathcal{A}_1}^\top
\widetilde{\mathbf{X}}_{\mathcal{A}_1}
\right)^{-1}
\right),
\end{align*}
which, as a function of $\nu_{11}$, is
\begin{align*}
&p(\widetilde{\boldsymbol{\beta}}_{\mathcal{A}_1} | \nu_{11}, \boldsymbol{\gamma}_1) \\
&\propto
\nu_{11}^{-(p_{\mathcal{A}_1}+1)/2}
\exp\left(
-\frac{1}{2g_1\nu_{11}}
\widetilde{\boldsymbol{\beta}}_{\mathcal{A}_1}^\top
\widetilde{\mathbf{X}}_{\mathcal{A}_1}^\top
\widetilde{\mathbf{X}}_{\mathcal{A}_1}
\widetilde{\boldsymbol{\beta}}_{\mathcal{A}_1}
\right),
\end{align*}
where
\begin{align*}
p_{\boldsymbol{\gamma}_1} = \sum_{j=1}^p \gamma_{1j}.
\end{align*}

For conjugacy, assume
\begin{align*}
\nu_{11} \sim \mathrm{IG}(a_\nu,b_\nu),
\end{align*}
with density proportional to
\begin{align*}
p(\nu_{11})
\propto
\nu_{11}^{-(a_\nu+1)}
\exp\left(
-\frac{b_\nu}{\nu_{11}}
\right).
\end{align*}
Combining likelihood and prior terms gives
\begin{align*}
p(\nu_{11} | \text{rest})
&\propto
\nu_{11}^{-\left(a_\nu + \frac{n+p_{\boldsymbol{\gamma}_1}+1}{2} + 1\right)}
\exp\left(
-\frac{b_n}{\nu_{11}}
\right),
\end{align*}
where
\begin{align*}
b_n
&=
b_\nu
+
\frac{1}{2}
\left(
\mathbf{y}_1
-
\widetilde{\mathbf{X}}_{\mathcal{A}_1}
\widetilde{\boldsymbol{\beta}}_{\mathcal{A}_1}
\right)^\top
\left(
\mathbf{y}_1
-
\widetilde{\mathbf{X}}_{\mathcal{A}_1}
\widetilde{\boldsymbol{\beta}}_{\mathcal{A}_1}
\right)
\\
&\quad +
\frac{1}{2g_1}
\widetilde{\boldsymbol{\beta}}_{\mathcal{A}_1}^\top
\widetilde{\mathbf{X}}_{\mathcal{A}_1}^\top
\widetilde{\mathbf{X}}_{\mathcal{A}_1}
\widetilde{\boldsymbol{\beta}}_{\mathcal{A}_1}.
\end{align*}
Thus, the full-conditional posterior of $\nu_{11}$ is
\begin{align*}
\nu_{11} | \text{rest}
\sim
\mathrm{IG}(a_n,b_n),
\end{align*}
with
\begin{align*}
a_n
=
a_\nu + \frac{n+p_{\boldsymbol{\gamma}_1}+1}{2}.
\end{align*}

\subsection{Step 9: Sampling $\boldsymbol{\xi} | \text{rest}$}

Recall that the ordinal response is linked to the latent variable through
\begin{align*}
y_{2i} = k
\qquad \Longleftrightarrow \qquad
\xi_{k-1} < y_{2i}^* \le \xi_k,
\qquad
k = 1,\dots,K,
\end{align*}
with
\begin{align*}
-\infty = \xi_0 < \xi_1 < \cdots < \xi_{K-1} < \xi_K = \infty.
\end{align*}
Following \cite{albert1993bayesian}, inference for the threshold parameters $\boldsymbol{\xi}$ is based on this latent-variable representation.

From the posterior in \eqref{eq:posterior_all_no_VS}, the full conditional of $\boldsymbol{\xi}$ is
\begin{align*}
p(\boldsymbol{\xi} | \text{rest})
&\propto
\prod_{i=1}^n
\mathbf{1}\{\xi_{y_{2i}-1} < y_{2i}^* \le \xi_{y_{2i}}\}
\, p(\boldsymbol{\xi}),
\end{align*}
where $p(\boldsymbol{\xi})$ is taken to be constant over the ordered region
\begin{align*}
\{\xi_1 < \cdots < \xi_{K-1}\}.
\end{align*}
To derive the full conditional of a given interior cut-point $\xi_j$, with $j=1,\dots,K-1$, we only retain the indicator terms involving $\xi_j$. These come from:
\begin{itemize}
    \item observations with $y_{2i}=j$, for which $\xi_j$ acts as an upper bound, and
    \item observations with $y_{2i}=j+1$, for which $\xi_j$ acts as a lower bound.
\end{itemize}
Hence,
\begin{align*}
p(\xi_j | \text{rest})
&\propto
\prod_{i:y_{2i}=j} \mathbf{1}\{y_{2i}^* \le \xi_j\}
\prod_{i:y_{2i}=j+1} \mathbf{1}\{\xi_j < y_{2i}^*\}
\mathbf{1}\{\xi_{j-1} < \xi_j < \xi_{j+1}\},
\end{align*}
which can be rewritten as
\begin{align*}
p(\xi_j | \text{rest})
&\propto
\mathbf{1}\left\{
\max\left(
\xi_{j-1},
\max_{i:y_{2i}=j} y_{2i}^*
\right)
<
\xi_j
<
\min\left(
\xi_{j+1},
\min_{i:y_{2i}=j+1} y_{2i}^*
\right)
\right\}.
\end{align*}
Therefore, the full conditional is
\begin{align*}
\xi_j | \text{rest}
\sim
\mathrm{Uniform}\left(
\max\left(
\xi_{j-1},
\max_{i:y_{2i}=j} y_{2i}^*
\right),
\;
\min\left(
\xi_{j+1},
\min_{i:y_{2i}=j+1} y_{2i}^*
\right)
\right),
\qquad
j=1,\dots,K-1.
\end{align*}

To summarise, each interior cut-point is updated from a uniform distribution over the interval determined by the ordering constraint and the neighbouring latent observations. In the implementation, if one of the adjacent categories $j$ or $j+1$ has no observations, the update for $\xi_j$ is skipped in order to preserve a feasible ordered configuration.

\clearpage

\section{Simulation study comparing the proposed method with Bayesian variable selection for Gaussian copula regression}\label{supp:simulation_study}
\subsection{Simulation setting}

The objective of this simulation study is to benchmark the proposed partially collapsed Gibbs sampler (PCGS) against the Bayesian variable selection approach for general Gaussian copula regression (BVS4GCR) in \citet{alexopoulos2021bayesian}. Although the joint Gaussian copula formulation in \citet{alexopoulos2021bayesian} and the proposed conditional formulation are closely related, the comparison is not between two implementations of the same Bayesian model. The two approaches employ different prior specifications; for example, BVS4GCR places shrinkage priors on the precision matrix, whereas the proposed method specifies priors directly on the conditional-model parameters. In addition, different Markov chain Monte Carlo algorithms are used. Consequently, identical posterior inference is not expected, despite the close relationship between the two model formulations. Differences in variable-selection performance and computational efficiency should therefore be interpreted as reflecting both the prior specification and the sampling algorithm.

To ensure that the comparison reflects the motivating application, we consider a simulation setting based on the LSAC application. We generate $R=100$ independent datasets with $n=3{,}000$ observations and $p=20$ predictors, of which $q^*=5$ are active (non-zero) for each response. The true regression coefficient vectors are chosen to have partially overlapping support, allowing us to assess the ability of the competing methods to correctly identify response-specific and shared effects. Specifically, the active predictors are $\{1,\ldots,5\}$ for the continuous response and $\{1,2,3,6,7\}$ for the ordinal response.

We induce collinearity among the predictors following \citet{george1993variable} by setting
\[
\mathbf{x}_j=\mathbf{x}_j^*+\mathbf{z},
\qquad
\mathbf{x}_j^*,\mathbf{z}\sim N(0,1),
\]
resulting in an approximate pairwise correlation of $0.5$. We set $\delta_{12}=0.6$ and $\nu_{11}=0.5$. For each simulated dataset, both competing algorithms are run for $M=40{,}000$ Markov chain Monte Carlo (MCMC) iterations following a burn-in period of $4{,}000$ iterations.

We compare the competing methods across four scenarios corresponding to decreasing signal strengths. For response $r=1,2$, define the signal-to-noise ratio (SNR) as
\begin{equation}
\mathrm{SNR}_r
=
\frac{\mathrm{Var}(\mathbf{X}\boldsymbol{\beta}_r)}
{\mathrm{Var}(\mathrm{Noise}_r)}
=
\frac{\boldsymbol{\beta}_r^\top\boldsymbol{\Sigma}\boldsymbol{\beta}_r}
{\mathrm{Var}(\mathrm{Noise}_r)},
\end{equation}
where $\boldsymbol{\Sigma}=\mathrm{Var}(\mathbf{x}_i)$ denotes the covariance matrix of the predictors, and the marginal noise variances are
\[
\mathrm{Var}(\mathrm{Noise}_1)=\nu_{11}+\delta_{12}^2,
\qquad
\mathrm{Var}(\mathrm{Noise}_2)=1,
\]
obtained from \eqref{eq:joint_representation}. Under this construction, $\boldsymbol{\Sigma}$ has unit diagonal elements and off-diagonal elements equal to $0.5$. The active coefficients in Scenarios~2--4 are obtained by uniformly scaling those in Scenario~1. Specifically,

\begin{enumerate}
\item \textbf{Full-strength signal.} The active coefficients are
\[
\boldsymbol{\beta}_1=(0.5,-0.5,0.8,0.4,-0.6)^\top,
\qquad
\boldsymbol{\beta}_2=(0.7,0.5,-0.4,0.6,-0.3)^\top,
\]
yielding SNRs of approximately $1.17$ and $1.28$ for the continuous and ordinal responses, respectively.

\item \textbf{Half-strength signal.} The active coefficients from Scenario~1 are multiplied by $0.5$, reducing the SNRs to $0.29$ and $0.33$ for the continuous and ordinal responses, respectively.

\item \textbf{Quarter-strength signal.} The active coefficients from Scenario~1 are multiplied by $0.25$, reducing the SNRs to $0.07$ and $0.08$ for the continuous and ordinal responses, respectively.

\item \textbf{Tenth-strength signal.} The active coefficients from Scenario~1 are multiplied by $0.10$, reducing the SNRs to $0.012$ and $0.013$ for the continuous and ordinal responses, respectively.
\end{enumerate}

Variable selection is based on the median probability model of \citet{BarbieriBerger2004}, in which the selected model consists of all predictors satisfying
\[
\Pr(\gamma_j=1 | \mathbf{y}) \ge 0.5.
\]

\subsection{Variable-selection comparison}

Variable-selection performance is assessed using sensitivity, specificity, false discovery rate (FDR) and the F1 score. Sensitivity is the proportion of truly active predictors that are correctly identified, while specificity is the proportion of truly inactive predictors that are correctly excluded. The FDR is the proportion of selected predictors that are false positives, whereas the F1 score is the harmonic mean of precision and sensitivity, providing a single summary measure that balances false positives and false negatives.

Table~\ref{tab:comparison-bvs4gcr} summarises the variable-selection performance of PCGS and BVS4GCR across the four signal-strength scenarios for the continuous ($y_1$) and ordinal ($y_2$) responses, while Table~\ref{tab:paired-diff} reports the corresponding paired differences across the $R=100$ simulated datasets together with 95\% confidence intervals. Under the full- and half-strength signal settings, both methods perform almost identically, achieving near-perfect sensitivity, specificity and F1 scores for both responses. The paired differences in Table~\ref{tab:paired-diff} confirm that the observed differences are negligible in these settings, with confidence intervals centred close to zero. Differences become apparent as the signal strength decreases. In the quarter-strength scenario, the proposed PCGS maintains almost perfect performance for the continuous response and substantially higher sensitivity for the ordinal response than BVS4GCR, while incurring only a slight reduction in specificity. The paired differences in Table~\ref{tab:paired-diff} confirm that the improvements in sensitivity and F1 score for the ordinal response are substantial, with confidence intervals excluding zero. Under the most challenging tenth-strength scenario, this trend becomes even more pronounced. The proposed PCGS exhibits considerably higher sensitivity and F1 scores for both responses, while sacrificing only a small amount of specificity. The paired differences in Table~\ref{tab:paired-diff} again confirm that these improvements are statistically significant. Overall, the proposed PCGS provides a more favourable trade-off between false positives and false negatives, particularly in low signal-to-noise settings where accurate recovery of active predictors is most challenging.

\begin{table}[H]
\centering
\caption{Comparison of variable-selection performance between the proposed partially collapsed Gibbs sampler (PCGS) and the Bayesian variable-selection approach for Gaussian copula regression (BVS4GCR) across four signal-strength scenarios. Results are averaged over $R=100$ simulated datasets with $n=3{,}000$ observations, $p=20$ predictors and $q^*=5$ active predictors for each response. Higher sensitivity, specificity and F1 scores indicate better variable-selection performance, whereas lower false discovery rates (FDR) are preferred.}
\label{tab:comparison-bvs4gcr}
\small
\begin{tabular}{lllcccc}
\toprule
Setting & Outcome & Method & Sensitivity & Specificity & FDR & F1 \\
\midrule

Full-strength
& $y_1$ (continuous) & PCGS     & 1.000 & 0.994 & 0.015 & 0.992 \\
&                     & BVS4GCR & 1.000 & 0.999 & 0.003 & 0.999 \\
& $y_2$ (ordinal)    & PCGS     & 1.000 & 0.998 & 0.005 & 0.997 \\
&                     & BVS4GCR & 1.000 & 1.000 & 0.000 & 1.000 \\

\addlinespace

Half-strength
& $y_1$ (continuous) & PCGS     & 1.000 & 0.997 & 0.007 & 0.996 \\
&                     & BVS4GCR & 1.000 & 0.999 & 0.003 & 0.998 \\
& $y_2$ (ordinal)    & PCGS     & 1.000 & 0.996 & 0.010 & 0.995 \\
&                     & BVS4GCR & 0.998 & 1.000 & 0.000 & 0.999 \\

\addlinespace

Quarter-strength
& $y_1$ (continuous) & PCGS     & 0.998 & 0.993 & 0.016 & 0.990 \\
&                     & BVS4GCR & 0.990 & 0.995 & 0.012 & 0.988 \\
& $y_2$ (ordinal)    & PCGS     & 0.886 & 0.997 & 0.011 & 0.928 \\
&                     & BVS4GCR & 0.516 & 1.000 & 0.000 & 0.666 \\

\addlinespace

Tenth-strength
& $y_1$ (continuous) & PCGS     & 0.458 & 0.995 & 0.028 & 0.602 \\
&                     & BVS4GCR & 0.286 & 1.000 & 0.000 & 0.427 \\
& $y_2$ (ordinal)    & PCGS     & 0.232 & 0.997 & 0.028 & 0.359 \\
&                     & BVS4GCR & 0.098 & 0.999 & 0.010 & 0.163 \\

\bottomrule
\end{tabular}
\end{table}

\begin{table}[H]
\centering
\caption{Paired differences in variable-selection performance between the proposed partially collapsed Gibbs sampler (PCGS) and the Bayesian variable-selection approach for Gaussian copula regression (BVS4GCR), computed as PCGS minus BVS4GCR over $R=100$ simulated datasets. Entries report the mean paired difference, with the corresponding 95\% confidence interval shown below in square brackets. Outcomes $y_1$ and $y_2$ denote the continuous and ordinal responses, respectively. Positive paired differences indicate improved performance of PCGS for sensitivity, specificity and F1 score, whereas negative paired differences indicate improved performance of PCGS for the false discovery rate (FDR).}
\label{tab:paired-diff}
\small
\begin{tabular}{llcccc}
\toprule
Setting & Outcome & $\Delta$ Sensitivity & $\Delta$ Specificity & $\Delta$ FDR & $\Delta$ F1 score\\
\midrule

Full-strength
& $y_1$
& \makecell[c]{0.000\\{\footnotesize [0.000, 0.000]}}
& \makecell[c]{-0.005\\{\footnotesize [-0.009, -0.001]}}
& \makecell[c]{0.012\\{\footnotesize [0.002, 0.022]}}
& \makecell[c]{-0.006\\{\footnotesize [-0.012, 0.000]}}\\

& $y_2$
& \makecell[c]{0.000\\{\footnotesize [0.000, 0.000]}}
& \makecell[c]{-0.002\\{\footnotesize [-0.004, 0.000]}}
& \makecell[c]{0.005\\{\footnotesize [-0.001, 0.011]}}
& \makecell[c]{-0.003\\{\footnotesize [-0.007, 0.001]}}\\

\addlinespace

Half-strength
& $y_1$
& \makecell[c]{0.000\\{\footnotesize [0.000, 0.000]}}
& \makecell[c]{-0.001\\{\footnotesize [-0.003, 0.001]}}
& \makecell[c]{0.003\\{\footnotesize [-0.003, 0.009]}}
& \makecell[c]{-0.002\\{\footnotesize [-0.006, 0.002]}}\\

& $y_2$
& \makecell[c]{0.002\\{\footnotesize [-0.002, 0.006]}}
& \makecell[c]{-0.004\\{\footnotesize [-0.008, 0.000]}}
& \makecell[c]{0.010\\{\footnotesize [0.002, 0.018]}}
& \makecell[c]{-0.004\\{\footnotesize [-0.008, 0.000]}}\\

\addlinespace

Quarter-strength
& $y_1$
& \makecell[c]{0.008\\{\footnotesize [-0.002, 0.018]}}
& \makecell[c]{-0.002\\{\footnotesize [-0.006, 0.002]}}
& \makecell[c]{0.004\\{\footnotesize [-0.004, 0.012]}}
& \makecell[c]{0.002\\{\footnotesize [-0.006, 0.010]}}\\

& $y_2$
& \makecell[c]{0.370\\{\footnotesize [0.333, 0.407]}}
& \makecell[c]{-0.003\\{\footnotesize [-0.005, -0.001]}}
& \makecell[c]{0.011\\{\footnotesize [0.001, 0.021]}}
& \makecell[c]{0.262\\{\footnotesize [0.233, 0.291]}}\\

\addlinespace

Tenth-strength
& $y_1$
& \makecell[c]{0.172\\{\footnotesize [0.137, 0.207]}}
& \makecell[c]{-0.005\\{\footnotesize [-0.009, -0.001]}}
& \makecell[c]{0.028\\{\footnotesize [0.004, 0.052]}}
& \makecell[c]{0.175\\{\footnotesize [0.140, 0.210]}}\\

& $y_2$
& \makecell[c]{0.134\\{\footnotesize [0.107, 0.161]}}
& \makecell[c]{-0.002\\{\footnotesize [-0.004, 0.000]}}
& \makecell[c]{0.018\\{\footnotesize [-0.006, 0.042]}}
& \makecell[c]{0.196\\{\footnotesize [0.159, 0.233]}}\\

\bottomrule
\end{tabular}
\end{table}

\subsection{Sampling efficiency comparison}
Sampling efficiency is evaluated using the effective sample size (ESS), defined as
\[
\mathrm{ESS}
=
\frac{M}
{1+2\sum_{k=1}^{\infty}\eta_k},
\]
where $M$ is the number of post-burn-in Markov chain Monte Carlo samples and $\eta_k$ denotes the lag-$k$ autocorrelation of the Markov chain.

Table~\ref{tab:ess-comparison} summarises the average effective sample sizes (ESS) over the $R=100$ simulated datasets for the tenth-strength signal scenario. Across all regression coefficients, the proposed partially collapsed Gibbs sampler produces effective sample sizes that are typically between five and ten times larger than those obtained using the sampler in \citet{alexopoulos2021bayesian} from the same number of post-burn-in Markov chain Monte Carlo iterations. This improvement is observed consistently across both active and inactive regression coefficients, as well as for the continuous and ordinal responses. Similar improvements are observed under the remaining signal-strength scenarios (results not shown).

In addition, PCGS is substantially faster than BVS4GCR in this simulation setting. The proposed method requires, on average, 2.66 minutes per simulated dataset, compared with 12.77 minutes for BVS4GCR, corresponding to an approximate 4.8-fold reduction in runtime. Since the two methods are implemented in different programming languages, this runtime comparison should be interpreted with some caution.

\begin{table}[H]
\centering
\caption{Comparison of effective sample sizes (ESS) for the proposed partially collapsed Gibbs sampler (PCGS) and the Bayesian variable-selection approach for Gaussian copula regression (BVS4GCR) in \citet{alexopoulos2021bayesian} under the tenth-strength signal scenario. Both methods were run for $40\,000$ post-burn-in Markov chain Monte Carlo (MCMC) iterations, and the reported ESS values are averaged over the $R=100$ simulated datasets. Active predictors correspond to the true non-zero regression coefficients in the data-generating process. Larger ESS values indicate more efficient MCMC sampling. The ratio is computed as $\mathrm{ESS}_{\mathrm{PCGS}}/\mathrm{ESS}_{\mathrm{BVS4GCR}}$. Row $j$ reports results for coefficients $\beta_{1j}$ and $\beta_{2j}$ corresponding to the continuous ($y_1$) and ordinal ($y_2$) responses, respectively.}
\label{tab:ess-comparison}
\small
\begin{tabular}{crrrrcrrr}
\toprule
&
\multicolumn{4}{c}{$y_1$}
&
\multicolumn{4}{c}{$y_2$}
\\
\cmidrule(lr){2-5}\cmidrule(lr){6-9}
$j$
& Active
& $\mathrm{ESS}_{\mathrm{BVS4GCR}}$
& $\mathrm{ESS}_{\mathrm{PCGS}}$
& Ratio
& Active
& $\mathrm{ESS}_{\mathrm{BVS4GCR}}$
& $\mathrm{ESS}_{\mathrm{PCGS}}$
& Ratio
\\
\midrule
1  & $\checkmark$ & 3666 & 15385 & 4.20 & $\checkmark$ & 3388 & 11799 & 3.48\\
2  & $\checkmark$ & 2010 & 8135  & 4.05 & $\checkmark$ & 1539 & 6954  & 4.52\\
3  & $\checkmark$ & 1862 & 19569 & 10.51 & $\checkmark$ & 6417 & 15888 & 2.48\\
4  & $\checkmark$ & 3757 & 22953 & 6.11 &              & 8121 & 25073 & 3.09\\
5  & $\checkmark$ & 2532 & 18584 & 7.34 &              & 4374 & 17490 & 4.00\\
6  &              & 4992 & 24016 & 4.81 & $\checkmark$ & 2986 & 13667 & 4.58\\
7  &              & 6025 & 30067 & 4.99 & $\checkmark$ & 9175 & 28364 & 3.09\\
8  &              & 6552 & 34255 & 5.23 &              & 9175 & 31671 & 3.45\\
9  &              & 6158 & 33843 & 5.50 &              & 8748 & 31280 & 3.58\\
10 &              & 6561 & 33604 & 5.12 &              & 8679 & 30959 & 3.57\\
11 &              & 6917 & 33484 & 4.84 &              & 8855 & 31813 & 3.59\\
12 &              & 6397 & 33633 & 5.26 &              & 9073 & 31815 & 3.51\\
13 &              & 6805 & 33732 & 4.96 &              & 8891 & 31829 & 3.58\\
14 &              & 6719 & 33358 & 4.96 &              & 8819 & 31781 & 3.60\\
15 &              & 6520 & 32860 & 5.04 &              & 8773 & 30303 & 3.45\\
16 &              & 6658 & 33728 & 5.07 &              & 8166 & 30618 & 3.75\\
17 &              & 6640 & 33174 & 5.00 &              & 9073 & 31795 & 3.50\\
18 &              & 6366 & 33539 & 5.27 &              & 9228 & 31397 & 3.40\\
19 &              & 6579 & 33424 & 5.08 &              & 9113 & 31707 & 3.48\\
20 &              & 6386 & 33063 & 5.18 &              & 9109 & 31057 & 3.41\\
\bottomrule
\end{tabular}
\end{table}

\subsection{Conclusion}

The simulation results demonstrate that the proposed partially collapsed Gibbs sampler provides competitive and, under weaker signal settings, superior variable-selection performance compared with the Bayesian variable-selection approach for Gaussian copula regression (BVS4GCR) in \citet{alexopoulos2021bayesian}. The improvement is primarily driven by higher sensitivity and F1 scores while maintaining comparable specificity. In addition, the proposed sampler produces substantially larger effective sample sizes and considerably lower computational times, reflecting the computational advantages obtained through the conditional representation, analytic marginalisation and the partially collapsed Gibbs sampling scheme.

These gains should be interpreted in the context of the respective modelling frameworks. The BVS4GCR methodology is substantially more general, accommodating arbitrary combinations of response types through the Gaussian copula construction. In contrast, the proposed approach is tailored to the important special case of a continuous and an ordinal response. This additional structure permits an explicit conditional representation of the joint model and, consequently, analytic marginalisation of the latent continuous response. The resulting simplification leads to a more efficient Markov chain Monte Carlo algorithm while retaining all modelling features required for the LSAC application that motivated this work.

\end{document}